\documentclass[aps,onecolumn,11pt,floatfix,altaffilletter,superscriptaddress,
               preprintnumbers,tightenlines,showpacs,showkeys,notitlepage,nofootinbib,prd]{revtex4-2}
\usepackage{appendix}
\usepackage{amsmath}
\usepackage{amsfonts}
\usepackage{bbold}
\usepackage{bm}
\usepackage{hyperref}
\usepackage{graphicx}
\usepackage{stackengine}
\usepackage{color}
\usepackage{subfigure}

\usepackage{helvet}
\usepackage{times}
\usepackage{feynmp-auto}
\usepackage[capitalise]{cleveref}
\usepackage{siunitx}

\usepackage[inline]{enumitem}
\usepackage{orcidlink}
\usepackage{lipsum}
\usepackage{comment}
\usepackage[normalem]{ulem}
\usepackage{multirow}

\usepackage{makecell}

\definecolor{lime}{HTML}{A6CE39}
\DeclareRobustCommand{\orcidicon}{%
	\begin{tikzpicture}
	\draw[lime, fill=lime] (0,0) 
	circle [radius=0.16] 
	node[white] {{\fontfamily{qag}\selectfont \tiny ID}};
	\draw[white, fill=white] (-0.0625,0.095) 
	circle [radius=0.007];
	\end{tikzpicture}
	\hspace{-2mm}
}

\foreach \x in {A, ..., Z}{\expandafter\xdef\csname orcid\x\endcsname{\noexpand\href{https://orcid.org/\csname orcidauthor\x\endcsname}
	{\noexpand\orcidicon}}
}

\definecolor{OliveGreen}{cmyk}{0.64,0,0.95,0.40}

\newcommand{\sm}{\text{\,SM}}
\newcommand{\nsi}{\text{\,NSI}}
\newcommand{\tr}{\text{\,true}}
\newcommand{\reco}{\text{\,reco}}
\newcommand{\n}{\text{\,Near}}
\newcommand{\f}{\text{\,Far}}
\newcommand{\bins}{n}
\newcommand{\mt}{{\!\mu\tau}}
\newcommand{\epmt}{\epsilon_{\mu\tau}}
\newcommand{\funchi}{\chi_{\text{BC}}^2}
\newcommand{\Oscmt}{\nu_\mu \rightarrow \nu_\tau}
\newcommand{\NtrSM}{\mathcal{N}}
\newcommand{\PmtSM}{P_{\!\mu\tau}}
\newcommand{\NtrNSI}{\hat{\mathcal{N}}}
\newcommand{\PmtNSI}{\hat{P}_{\!\mu\tau}}
\newcommand{\lt}{\left(}
\newcommand{\rt}{\right)}
\newcommand{\sbin}{\sum_{i}^{\bins}}

\definecolor{lime}{HTML}{A6CE39}
\DeclareRobustCommand{\orcidicon}{%
	\begin{tikzpicture}
	\draw[lime, fill=lime] (0,0) 
	circle [radius=0.16] 
	node[white] {{\fontfamily{qag}\selectfont \tiny ID}};
	\draw[white, fill=white] (-0.0625,0.095) 
	circle [radius=0.007];
	\end{tikzpicture}
	\hspace{-2mm}
}

\foreach \x in {A, ..., Z}{\expandafter\xdef\csname orcid\x\endcsname{\noexpand\href{https://orcid.org/\csname orcidauthor\x\endcsname}
	{\noexpand\orcidicon}}
}

\definecolor{OliveGreen}{cmyk}{0.64,0,0.95,0.40}

\usepackage[utf8]{inputenc}
\usepackage{amsmath}
\usepackage{amsfonts}
\usepackage{amssymb}
\usepackage{calrsfs}
\usepackage[left=2cm,right=2cm,top=2cm,bottom=2cm]{geometry}
\usepackage[mathscr]{euscript}

\begin{document}

\title{Tau neutrino as a probe of charged nonstandard interactions at DUNE}

\author{A. Cherchiglia }
 
\author{O. L. G. Peres }
 
\author{E. S. Souza }
    \affiliation{Instituto de F\'isica Gleb Wataghin, Universidade Estadual de Campinas,\hspace{0.5cm}Rua Sérgio Buarque de Holanda, 777, 13083-859, Campinas, SP, Brazil}


\begin{abstract}
   In this work, we study the influence of charged-current non-standard interactions (CC-NSI) at a DUNE-like experiment. We are particularly interested in the tau neutrino channel accessible at DUNE, given the higher energy neutrino flux expected to be achieved by the experiment. We focus on CC-NSI that may affect the pion decay, the primary source of neutrinos for DUNE. We show that the expected sensitivity at the near detector may supersede the present bounds coming directly from pion decay by one order of magnitude.
   \end{abstract}
\maketitle
%


\section{Introduction}

The Long-Baseline Neutrino Facility to Deep Underground Neutrino Experiment (LBNF-DUNE) is one of the long baseline next generation neutrino oscillation experiments, whose main goals include detecting the occurrence of CP violation in the leptonic sector, obtaining the octant of the parameter $\theta_{23}$ together with the mass ordering. Given the precision to be achieved by the next-generation experiments, the three-neutrino paradigm will be further scrutinized. In particular, the expected accuracy may allow us to constrain some effects of new physics emerging through neutrino oscillations at the per cent level~\cite{DUNE:2020ypp,DUNE:2024wvj}. Several studies have been proposed focusing on the sensitivity of the next-generation experiments to different new-physics scenarios (see, for instance,~\cite{Arguelles:2022tki} for a recent review). However, their emphasis is mainly on the exploration of the muon neutrino $\nu_\mu$ disappearance and electron neutrino $\nu_{\rm e}$ appearance channels (plus antineutrinos). Given the unprecedented energy of the neutrino flux at DUNE, it will also be possible to investigate the tau neutrino $\nu_\tau$ appearance channel. Even more interestingly, the collaboration plans to perform a dedicated high-energy run, which would significantly increase the number of $\nu_\tau$ events~\cite{DUNE:2020ypp,DUNE:2024wvj}. Given this panorama, it may be possible to constrain new physics scenarios in connection with tau neutrinos as well~\cite{Giunti:2001ws,Bertou:2001vm,DeGouvea:2019kea,Farzan:2021gbx,Saveliev:2021jtw,Martinez-Soler:2021sir,Denton:2021rsa,Ansarifard:2021dju,Huang:2021mki,MammenAbraham:2022xoc,Huang:2022ebg,Bakhti:2023mvo,Meighen-Berger:2023xpr,Dev:2023rqb,IceCube:2024nhk,Francener:2024ney,Francener:2024euo,Huege:2024nic,IceCube:2024nhk,KM3NeT:2025ftj,Yu:2025cyx}. However, there are few works that focus on new physics scenarios with impact at $\nu_\tau$ events at DUNE~\cite{DeGouvea:2019kea,Ghoshal:2019pab,Machado:2020yxl,Dev:2023rqb,Bakhti:2023mvo,Yu:2025cyx,DUNE:2024wvj}. In this work, our aim is to shed light on new physics scenarios affecting $\nu_\tau$ events through modifications at the pion decay rates. We recall that for long-baseline experiments, the primary production source of neutrinos is from the pion decay; thus, any modification may impact the observed events at the near/far detector.

In general, new physics effects may be parametrized from a UV completion or a purely effective perspective. In the context of neutrino experiments, it is customary to follow the second route,  parametrizing those effects in terms of Non-Standard Interactions (NSI), which represent new sources for charged (neutral) currents apart from the usual W (Z) mediated processes; see~\cite{Farzan:2017xzy,Proceedings:2019qno,Arguelles:2022tki} for reviews. The connection between NSI and the Standard Model effective field theory (SMEFT) was performed some time ago~\cite{Bischer:2019ttk}, while the translation to the Low energy EFT (LEFT) was analysed in a series of works, for instance, \cite{Bergmann:1999pk,Antusch:2008tz,Gavela:2008ra,Meloni:2009cg,Altmannshofer:2018xyo,Falkowski:2019kfn,Falkowski:2019xoe, Babu:2019mfe,Bischer:2019ttk,Davidson:2019iqh,Terol-Calvo:2019vck,Babu:2020nna,Du:2020dwr,Falkowski:2021bkq,Du:2021rdg,Breso-Pla:2023tnz,Kopp:2025ffx}. Regarding UV completions, it is possible to map all heavy (compared to the electroweak scale) weakly coupled models based on the SM gauge group that can produce NSI, at tree-level matching. The map is achieved by resorting to the tree-level SMEFT dictionary~\cite{deBlas:2017xtg}. In~\cite{Cherchiglia:2023aqp}, it was shown how to automatically implement this idea, allowing one to identify the connection between NSI and particular UV completions with particles with masses at the TeV scale or above. For the present work, we will be interested in scenarios that affect the pion decay, which can be parametrized by non-standard interactions from charged-currents (CC-NSI). In particular, those can be generated by models containing leptoquarks or extra scalar doublets at tree-level matching~\cite{Cherchiglia:2023aqp}. Taking into account the one-loop matching, the list of possible models is much larger, found in~\cite{Cherchiglia:2025ufn}. However, in this work, we will not delve into UV completion but rather consider a pure EFT perspective. Therefore, our main aim is to find the sensitivity from DUNE at CC-NSI related to the tau neutrino appearance. 

Our work is organized as follows: in section~\ref{sec:eft} we provide details of the EFT approach for CC-NSI, and set our notation. Section~\ref{sec:sample} is devoted to the description of the simulation for a DUNE-like experiment, regarding the tau neutrino appearance channels. In section~\ref{sec:constraints} we describe constraints from other experiments or channels, while in section~\ref{sec:numerical} we present our numerical analysis, in particular the sensitivity for CC-NSI. We conclude in section~\ref{sec:conclusion}.


\section{EFT description for CC NSI}
\label{sec:eft}

It is customary to parametrize new effects due to the interaction of neutrinos with a medium as non-standard interactions (NSI). In this work, we adopt a more general approach, considering NSI to denote not only BSM effects (new interactions) at neutrino propagation, but also at its production or detection. To make this distinction more straightforward, we notice that NSI connected with production are connected with charged-current (CC) processes (for the experiments to be analysed in this work, the pion decay), while NSI connected with propagation are related to neutral currents (NC) processes. Restricting our attention to CC-NSI, they can be conveniently expressed by the EFT Lagrangian below~\cite{Falkowski:2018dmy,Falkowski:2019kfn,Falkowski:2021bkq,Falkowski:2021bkq,Kopp:2024yvh,Kopp:2025ffx}

\begin{align}
    \mathcal{L}_{\rm WEFT} 
\supset
- 
   \frac{2\, V_{jk}^{\rm CKM}}{v^2} 
   \Big\{
& 
   [1+\epsilon_L^{jk}]_{\alpha \beta} 
   \left( 
     \bar{u}^j\gamma^{\mu}P_L d^k
   \right)
   \left(
     \bar{\ell}_{\alpha}\gamma_{\mu}P_L \nu_{\beta} 
   \right)
+ 
  [\epsilon_R^{jk}]_{\alpha \beta} 
  \left( 
    \bar{u}^j\gamma^{\mu}P_R d^k
  \right)
  \left(
    \bar{\ell}_{\alpha}
    \gamma_{\mu}
    P_L 
    \nu_{\beta} 
  \right)
    \nonumber\\
& 
  \left. +\frac{1}{2}\left[ \epsilon_S^{jk}\right]_{\alpha \beta} \left(\bar{u}^jd^k \right)\left(\bar{\ell}_{\alpha}P_L \nu_{\beta} \right)
    -\frac{1}{2}\left[ \epsilon_P^{jk}\right]_{\alpha \beta} \left(\bar{u}^j \gamma^5 d^k \right)\left(\bar{\ell}_{\alpha}P_L \nu_{\beta} \right)\right.
    \nonumber\\
& 
   +  \frac{1}{4}\left[ \epsilon_T^{jk}\right]_{\alpha \beta} \left(\bar{u}^j \sigma^{\mu \nu} P_L d^k \right)\left(\bar{\ell}_{\alpha} \sigma_{\mu \nu}P_L \nu_{\beta} \right) +\mathrm{h.c.} 
    \Big\}~.
\label{eq:lagrangian}
\end{align}
Here, $V^{\rm CKM}$ is the Cabibbo-Kobayashi-Maskawa (CKM) matrix, $v = 
1/(\sqrt{2}G_F) \approx 246 $\,GeV is the vacuum expectation value 
of the SM Higgs field, $P_L$ is left chiral projector and $\epsilon_X$ are the Wilson coeficients, 
with $X~=~L, R, S, P, T$ for left-handed, right-handed, scalar, 
pseudo-scalar and tensor, respectively. The Roman (Greek) symbols 
denote the quark (lepton) generations. This Lagrangian is based on 
the Weak Effective Field Theory (WEFT), defined below the electroweak 
scale.

In this work, we will be concerned only with CC NSI related to pion decay; therefore, we need only to consider the case $j=k=1$, and, for simplicity, drop the overscript on the WC hereafter. Moreover, it was shown (for example in~\cite{Falkowski:2019kfn}) that only the WCs $\epsilon_{L,R,P}$ can affect the decay of the pion. Since the pseudoscalar coefficient involves a chirality flipping (while the standard vertex, with the W-boson, is chirality conserving), the WC $\epsilon_{P}$ will appear multiplied by a ratio of the pion mass over the mass of its constituents, as well as the emitted charged lepton. Therefore, we will be able to constrain $\epsilon_{P}$ much more than $\epsilon_{L,R}$, motivating us to only consider $\epsilon_{P}$ to be non-null. Moreover, for simplicity, we denote hereafter $(\epsilon_{P})_{\alpha\beta}$ as $\epsilon_{\alpha\beta}$.

Using a QFT description for neutrino oscillation, the differential event rate for flavor neutrinos $\beta$ with energy $E_\nu$ to be detected at a distance $L$ from the source $S$, where they were produced with flavor $\alpha$, is given by~\cite{Breso-Pla:2023tnz}  \begin{equation}\label{eq:rate:final}
    R_{\alpha\beta}^S = N_T\sigma_\beta^{\text{SM}}(E_{\nu})\Phi_\alpha^{\text{SM}}(E_{\nu})
    \sum_{k,l}
    e^{-i\frac{L \Delta m_{kl}^{2}}{2E_{\nu}}} \frac{[ {\cal P} U ]_{\alpha l}   [U^\dagger  {\cal P}^\dagger]_{k \alpha}} {[{\cal P}{\cal P}^\dagger]_{\alpha\alpha}} U_{\beta k}U_{\beta l}^{*}~,
\end{equation}
where $\Phi_\alpha^{\text{SM}}$ is the SM flux, $\sigma_\beta^{\text{SM}}$ is the SM cross-section, $U$ is the Pontecorvo–Maki–Nakagawa–Sakata (PMNS) matrix in matter~\cite{Maki:1962mu,Pontecorvo:1957cp} and ${\cal P}$ is due to the CC NSI, being given by
\begin{eqnarray}
     [ {\cal P}]_{\alpha \beta} 
     &\equiv& 
     \delta_{\alpha \beta}   
    -
    \frac{ m_{\pi^\pm}^2}{   m_{\ell_\alpha}  (m_u + m_d) } \epsilon_{\alpha \beta} \, .
\label{eq:TermNSI}
\end{eqnarray}

The denominator in eq.\eqref{eq:rate:final} is due to indirect effects, first introduced in~\cite{Breso-Pla:2023tnz}. It appears by writing the event rate in terms of the measured pion decay (thus defining $\Phi_\alpha^{\text{SM}}$), which is an inclusive measurement regarding the emitted neutrino. As discussed in~\cite{Cherchiglia:2023aqp}, if only diagonal CC NSI are considered, due to the indirect effects, they will cancel in the event rate. This is the case if a UV completion for the CC NSI comes only from BSM particles with masses above the electroweak scale. In the present work, we will not go into UV completions. However, since we will consider only non-diagonal CC NSI (to be more precise, $\epsilon_{\mu\tau}$), UV completion will typically require the inclusion of extra symmetries beyond the SM gauge group and the presence of BSM particles with masses below the electroweak scale.


\section{Tau neutrino sample at DUNE}
\label{sec:sample}

As stated in the introduction, few works are dedicated to investigating how new physics effects modify the $\nu_\tau$ appearance channel. We are particularly concerned with changes in the pion decay rate, which may even help alleviate the present tension between the T2K and NO$\nu$A experiments~\cite{Cherchiglia:2023ojf}. A simulation of the expected number of $\nu_\tau$ events in DUNE was performed in~\cite{DeGouvea:2019kea} using the Standard Oscillation Model (SOM). In this section, we discuss how to incorporate the NSI in the pion decay with knowledge of the $\nu_\tau$ appearance events simulated in~\cite{DeGouvea:2019kea}. We considered both near- and far detectors in our sensitive analyses. 

The DUNE experiment consists of two liquid argon time projection chamber (LArTPC) detectors: a near detector (ND) located 574~m from the neutrino source at Fermilab, and a far detector (FD) situated 1297~km away at the Sanford Underground Research Facility in South Dakota. The neutrino beam is produced at the Long Baseline Neutrino Facility (LBNF) by directing approximately $1.1 \times 10^{21}$ protons per year onto a fixed target. The ND, with a fiducial mass of 70~tons, is composed of multiple subdetectors and is used to measure the initial neutrino flux and interaction rates. The beam then travels through the Earth's crust, with an average matter density $\rho = 2.848~\mathrm{g/cm}^3$ and electron fraction $Y_e = 1/2$, before reaching the FD---a 40~kton (fiducial mass) detector employing LArTPC technology to reconstruct neutrino interactions with high precision.

The DUNE experiment is expected to be unique in the detection of $\nu_\tau$ events. Especially there is the proposal to run $\tau$-optimized fluxes denoted by high-energy mode, where a more significant fraction of $\nu_\tau$ is generated above the $\tau$-production threshold. In our simulation, we assumed to collect events for three years in both neutrino mode (forward horn current) and antineutrino mode (reverse horn current), plus one year of high-energy mode (forward horn current). This seven-year data collection will be indicated as $3\:+\: 3\: +\: 1$.  

Given the much higher flux of muon neutrinos, the oscillation parameters are determined in long-baseline experiments due to the appearance and disappearance of $\nu_{\rm e}$ and $\nu_{\mu}$ events in general. To classify the neutrinos for flavor, it is necessary to reconstruct both the outgoing charged leptons and the energy of the incoming neutrino. Identifying and reconstructing $\nu_\tau$ samples via a charged-current (CC) interaction is naturally more challenging than detecting and reconstructing $\nu_{\rm e}$ and $\nu_\mu$ events for a variety of reasons. Since the $\tau$ is the heaviest lepton, there is a large threshold on the energy of incoming $\nu_\tau$ to generate CC interaction with the detector's medium. For a CC interaction with nucleon ($\nu_\tau + N \rightarrow \tau + N'$) to happen, the energy of the incoming $\nu_\tau$ needs to fulfill $E_{\nu} \gtrsim 3.35~$~GeV. Furthermore, reconstructing the $\tau$-lepton in the neutrino detector is a greater challenge when compared to other charged leptons. Even 
though liquid argon detectors have a resolution of around several millimeters, the $\tau$-leptons decay promptly, with a decay length significantly smaller than the resolution of liquid argon detectors. Therefore, the identification and reconstruction of $\tau$-leptons occur by reconstruction of 
its decay products, which always includes a neutrino in the final-state $(\tau \rightarrow \nu_\tau + {\rm something \ 
else})$. Around 65\% of $\tau$-decays are hadronic~\cite{ParticleDataGroup:2024cfk}, which represents another challenge. All these facts make it more difficult to correctly identify a scattering event as a $\nu_\tau$ CC interaction, rejecting neutral-current (NC) background, and correctly reconstructing 
the neutrino of the incoming $\nu_\tau$. Despite all challenges, the combination of a large $\nu_\tau$ flux above the $\tau$-production energy threshold and the capability of the liquid argon detector in reconstructing the final states makes DUNE one of the best experiments for detecting $\nu_\tau$ samples. 
For the 3 + 3 + 1 scenario, the DUNE collaboration expects to collect more than 300 events~\cite{DeGouvea:2019kea}.
 
Regarding the neutrino flux at DUNE, the $\nu_\tau$ 
production in the Standard Model (SM) is basically 
insignificant and can be ignored. The $\nu_\tau$ samples will arise due to the oscillation $\Oscmt$ at the FD
\begin{eqnarray}
    \frac{
            d \NtrSM_{\mt}
        }{
            d E_\nu^\tr
        }
    =
    \sbin
    \lt
    \phi_{\mu}^\sm \:
    \sigma_{\tau}^\sm \:
    \PmtSM^\sm
    \rt_i ~ ,
\label{eq:General_EventSM_true}
\end{eqnarray}
where $\phi_\mu^{\sm}$ is the flux of $\nu_\mu$ from the source, $\sigma_\tau^{\sm}$ the cross section of $\nu_{\tau}$ in argon, and $\PmtSM^{\sm}\equiv P(\Oscmt)$ the usual oscillation probability. Note that the events in eq.~\ref{eq:General_EventSM_true}) are divided into energy bins, where $i\!=\!\{1,2,\cdots,\bins\}$. Instead of performing a full simulation for the neutrino flux and cross-section, we will read the number of true events from \cite{DeGouvea:2019kea}. It is possible to map the reconstructed events per energy bin ($E_{\nu}^{\reco}$) from the true events per energy bin ($E_{\nu}^{\tr}$) using a migration matrix (smeared)

\begin{align}
    \frac{dN_{\mu\tau}}{dE_{\nu}^{\reco}}
    =
    \int\! dE_{\nu}^{\tr}\: 
    \frac
    { 
        d \NtrSM_{\mt} 
    }
    {
        dE_{\nu}^{\tr}
    }\:
    f(E_\reco,E_\tr) ~ .
\label{eq:General_EventSM_reco}
\end{align}
In general, the mapping $f(E_\reco,E_\tr)$ is described using a Gaussian function which takes into account the detector's resolution as well as the bias and error in the reconstruction of the energy of incoming neutrinos. In other words, the quantity observed by detectors is the reconstructed events. We will adopt the following Gaussian function~\cite{DeGouvea:2019kea} 
\begin{eqnarray}
    f(E_\reco,E_\tr)
    =
    \exp
    \left[
        -\frac{1}{2}
        \left(
            \frac{
                E_{\nu}^{\reco}
                -
                \mu_{\nu}^{\tr}
                }
                {
                \sigma_{\nu}^{\tr}
                }
        \right)^2
    \right] ~ ,
\label{eq:Mapping_Matrix}
\end{eqnarray}
where $\mu_{\nu}^{\tr} = b E_{\nu}^{\tr}$ is the Gaussian mean value and $\sigma_{\nu}^{\tr} = r E_{\nu}^{\tr}$ its width, where $b$ and $r$ are the bias and resolution of detector, respectively. The best fit we found is $b = 0.435$ and $r = 0.255$. Recently, the NuFit 
collaboration~\cite{Esteban:2024eli} released an update of the oscillation parameters, which we quote below (for simplicity, we assume normal ordering hereafter)
\begin{eqnarray}
&
    \sin^2\theta_{12} = 0.307,
    \quad
    \sin^2\theta_{13} = 0.02195,
    \quad
    \sin^2\theta_{23} = 0.561,
    \quad
    \delta_{CP}/\pi = 0.98,
\nonumber 
\\ 
& 
    \Delta m_{21}^2 = 7.49 
    \times 10^{-5}~{\rm eV^2},
    \quad
    \Delta m_{31}^2 = +2.534
    \times 10^{-3}~ {\rm eV^2}.
\label{eq:NuFit}
\end{eqnarray}
Since we are using the true events from~\cite{DeGouvea:2019kea} that were computed using the 2018 global best fit values, we rescaled the simulated data considering the relation 
\begin{eqnarray}
    \NtrSM_{(24)}^{\tr}
    =
    \phi_{\mu}^{\sm} \: 
    \sigma_{\tau}^{\sm} \:
    \PmtSM^{(24)}
    =
    \frac{
            \PmtSM^{(24)}
        }
        {
            \PmtSM^{(18)}
        }
    \NtrSM_{(18)}^{\tr} ~ ,
\label{eq:EventSM_true}
\end{eqnarray}
where $\NtrSM_{(ij)}^{\tr}$ and $\PmtSM^{(ij)}$ represent the true events at FD eq.~(\ref{eq:General_EventSM_true}) and the 
oscillation probability $\Oscmt$, respectively, assuming the best-fits for the years 2018 or 2024 ( denotated $(ij) = (18) \ {\rm or} \ (24)$). To obtain the updated reconstructed events $\big(N_{(24)}^{\reco}\big)$, we applied the Gaussian mapping given in eq.~\ref{eq:Mapping_Matrix}) in the updated true events of eq.~(\ref{eq:EventSM_true}). We used the updated reconstructed events to carry out our analysis in section \ref{sec:numerical}. From now on, we omit the upper and lower indexes when there are no ambiguities.

The introduction of the CC NSI to the Lagrangian 
eq.~(\ref{eq:lagrangian}) results in new channels for pion decay~\cite{Cherchiglia:2023ojf}, where we assume only the pseudo-scalar interaction $[\epsilon_{P}]_{\mu\tau} \equiv \epmt$ is non-null. This new channel violates the lepton flavor number, the $\mu$-lepton can be associated with $\nu_\tau$ as a sub-leading channel decay because $\epmt\neq 0$. As discussed in~\cite{Cherchiglia:2023ojf}, and can be observed in the eq.~(\ref{eq:rate:final}), even though the CC NSI modifies the flux of neutrino flavor, we can summarize its effect by modifying the oscillation probability. The flux and cross section remain the same as in the SM. If only the term $\epmt$ is different from zero, the expression to NSI probability is very compact
\begin{align}
\PmtSM^{\nsi}
    =
    \left|
        S_{\tau \mu}^{\sm} 
        - 
        p_\mu 
        \epmt^{*}
        S_{\tau \tau}^{\sm}
    \right|^2 ~ ,
\label{eq:Probability_Smatrix}
\end{align}
where $S_{\alpha\beta}^{\sm}$ are the components $\alpha\beta = \{\tau\mu, \: \tau\tau\}$ of the evolution matrix in the SM, and $\epmt$ is the epsilon term from eq.~(\ref{eq:TermNSI}) for $\alpha\beta = \{\mu\tau\}$, with the associated coefficient represented by $p_{\mu} \equiv m_{\pi}^2/ \left(m_{\mu} (m_u + m_d) \right) \approx$ 27. For now on, we will adopt the notation for the NSI probability $\PmtSM^{\nsi} \equiv \hat{P}(\Oscmt) = \PmtNSI\!$, where we have used the hat on the NSI variables to differentiate them from the SM variables. 

Although we referred to 
eq.~(\ref{eq:Probability_Smatrix}) as probability, we emphasize that it can have values greater than one. The NSI probability with matter effect can be derived in a relatively simple perturbative analytical form~\cite{Cherchiglia:2023ojf}, by using perturbative expressions of the evolution matrix with matter effects~\cite{Asano:2011nj}. We have used these formulas to cross-check our final results, which were obtained without resorting to approximations. For simplicity, we will assume the $\epmt$ to be real. 

Before we present the discussion of how we generated the NSI events at FD, we should discuss how the knowledge of the events in the ND can be used to reduce the uncertainties at the FD 
\begin{eqnarray}
    N_{\!\mu\tau;\:\f}^{\text{\,Obs}}
    =
    \frac{
            N_{\!\mu\mu;\:\n}^{\text{\,Obs}} 
        }{
            N_{\!\mu\mu;\:\n}^{\text{\,Mod}}
        }
    \:
    N_{\!\mu\tau;\:\f}^{\text{\,Mod}} 
    =
    \frac{
            \hat{P}_{\!\mu\mu}^{\n}
        }{
            P_{\!\mu\mu}^{\n}
        }
    \:
    N_{\!\mu\tau;\:\f}^{\text{\,Mod}}
    \approx
    \hat{P}_{\!\mu\mu}^{\n}
    \:
    N_{\!\mu\tau;\:\f}^{\text{\,Mod}}
    ~ ,
\label{eq:Events_near_to_far}
\end{eqnarray}
where $N_{\alpha\beta;\:X}^Y$ represents the transition $\alpha\rightarrow\beta$ to the reconstructed events assuming the $X \!=\! \{$Near, Far$\}$ detector and the $Y \!=\! \{$Mod, Obs$\}$ case, the model's prediction and the observation, respectively. Since we aim to obtain the sensitivity to NSI, its effect should be present in the observation events. For completeness, we considered the relation eq.~(\ref{eq:Events_near_to_far}) in our analysis; however, the effect is irrelevant for this case. 

As we aim to obtain the observable events in the FD with the NSI effect included, we should consider the reconstructed NSI events after application of the migration matrix under the true NSI events. The true NSI events are obtained from the true SM events as follows
\begin{align}
    \frac{
            d \NtrNSI_\mt
        }{
            d E_\nu^\tr
        } 
    =
    \sbin
    \lt
    \phi^{\sm} \sigma^{\sm}
    \PmtNSI
    \rt_i
    =
    \frac{\PmtNSI}{\PmtSM}
    \frac{
            d \NtrSM_\mt
        }{
            d E_\nu^\tr
        } ~ ,
\label{eq:Events_true_NSI}
\end{align}
where $\NtrSM_\mt$ are the true events given in 
eq.~(\ref{eq:EventSM_true}) at the FD. Therefore, the reconstructed NSI events at the FD are obtained by applying the migration matrix 
eq.~(\ref{eq:Mapping_Matrix}) in 
eq.~(\ref{eq:Events_true_NSI}), analogous to 
eq.~(\ref{eq:General_EventSM_reco}), and substituting the end result into eq.~(\ref{eq:Events_near_to_far})
\begin{align}
    \frac{d\hat{N}_\mt}{dE_{\nu}^{\reco}}
    =
    \hat{P}_{\!\mu\mu}^{\n}
    \int\! dE_{\nu}^{\tr}\: 
    \frac
    { 
        d \NtrNSI_{\mt} 
    }
    {
        dE_{\nu}^{\tr}
    }\:
    f(E_\reco,E_\tr) ~ .
\label{eq:Events_reco_NSI}
\end{align}

\begin{figure}[!hbt]
  \centering
  \begin{minipage}{0.32\textwidth}
    \includegraphics[width=\linewidth]{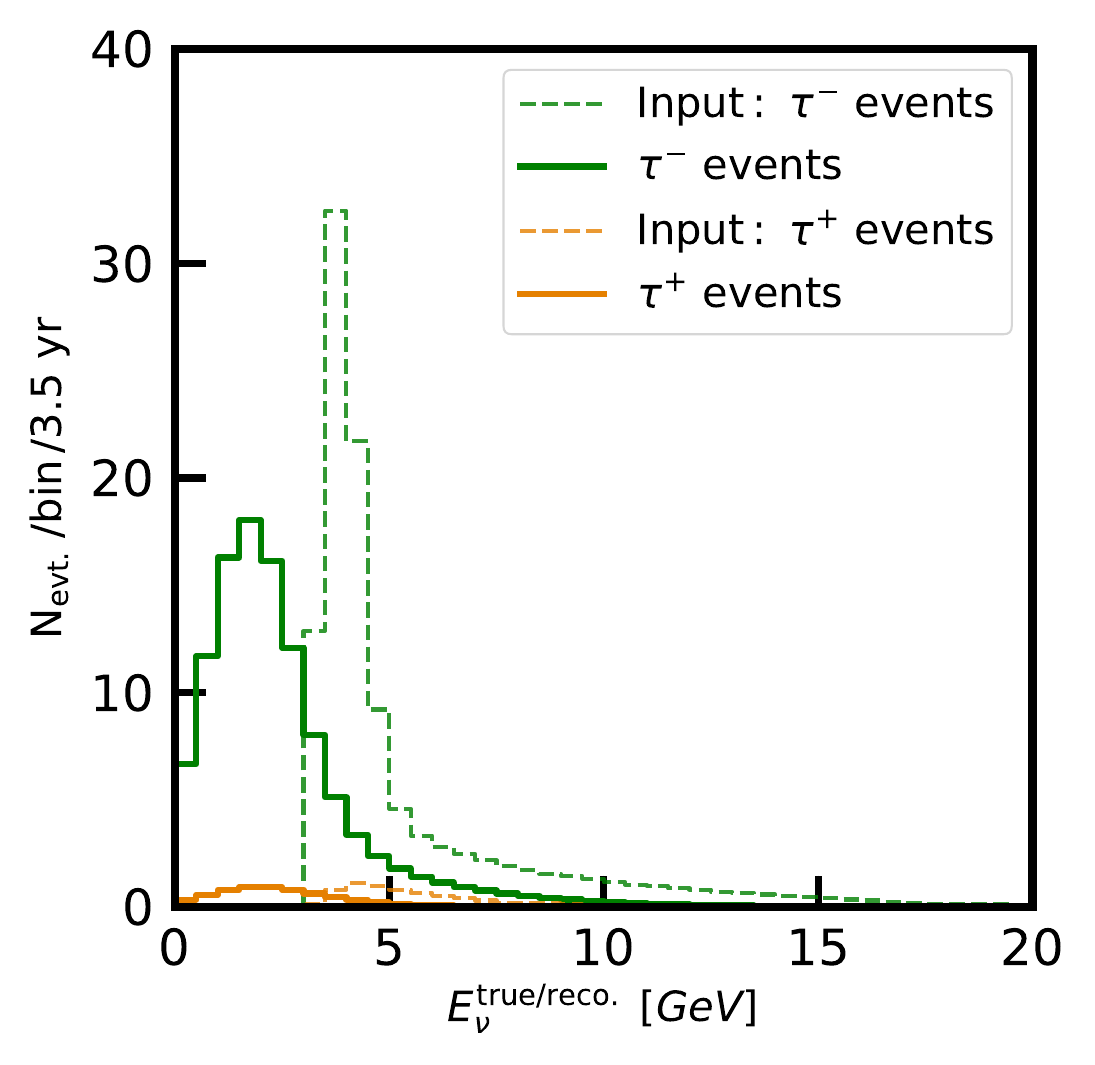}
  \end{minipage}\hfill
  \begin{minipage}{0.32\textwidth}
    \includegraphics[width=\linewidth]{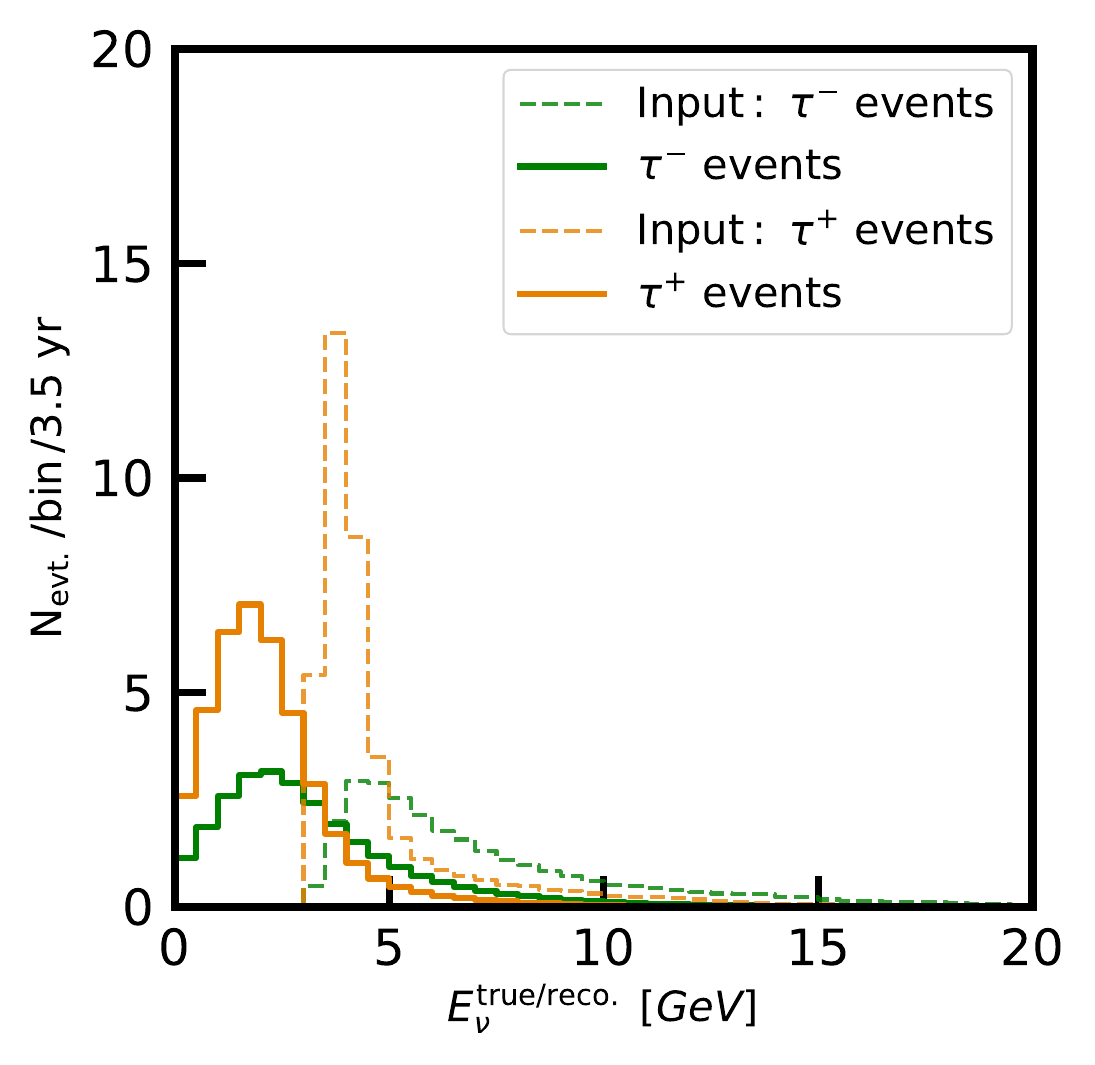}
  \end{minipage}\hfill
  \begin{minipage}{0.32\textwidth}
    \includegraphics[width=\linewidth]{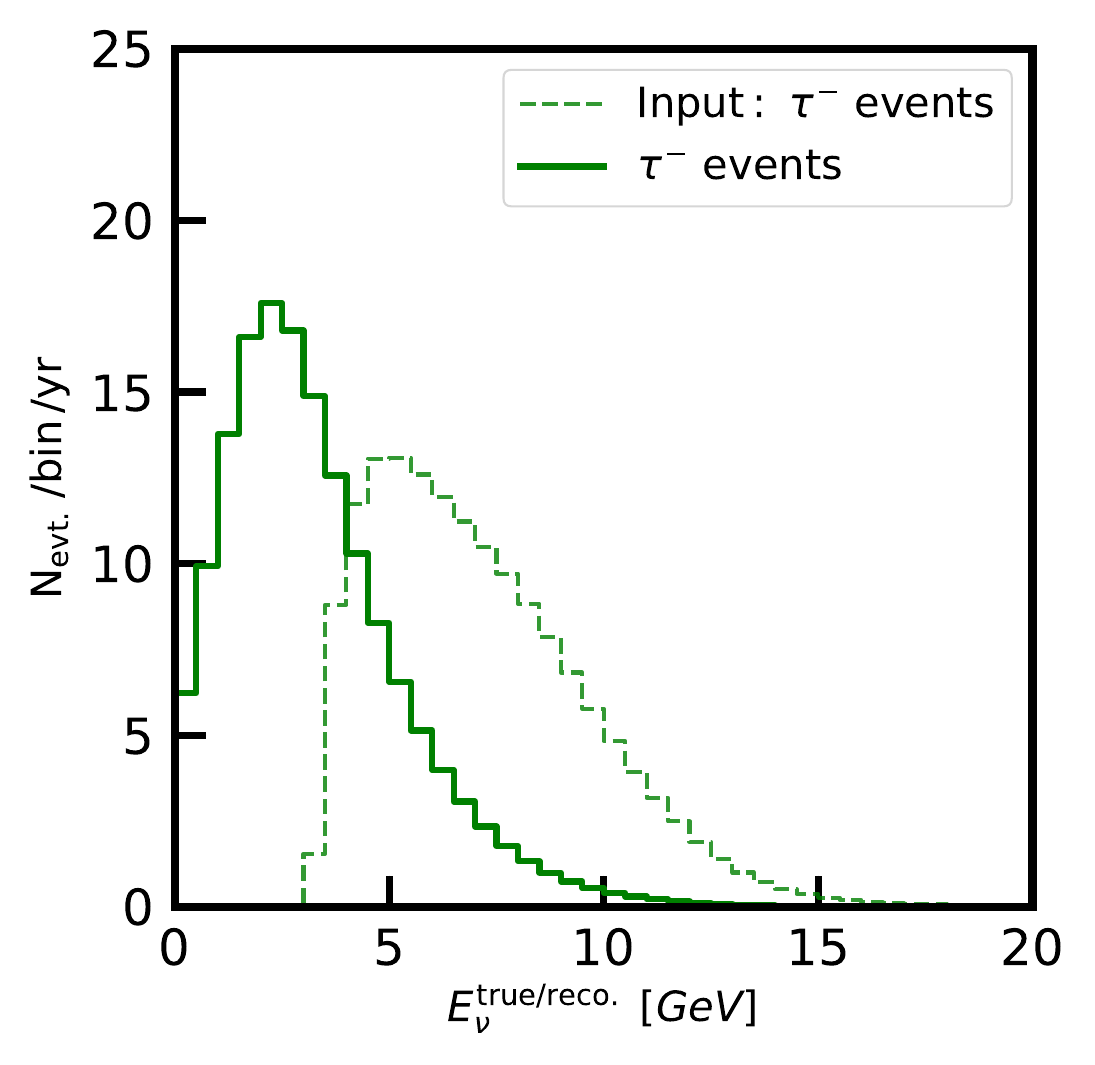}
  \end{minipage}
  \caption{Expected number of events after 3.5 years in neutrino mode (left), 3.5 years in antineutrino mode (middle), 1 year in high energy mode (right). Dashed lines represent the true number of events, while solid lines represent the reconstructed ones.}
  \label{fig:Plot_Event_SM_reco}
\end{figure}

\begin{figure}[!hbt]
  \centering
  \begin{minipage}{0.32\textwidth}
    \includegraphics[width=\linewidth]{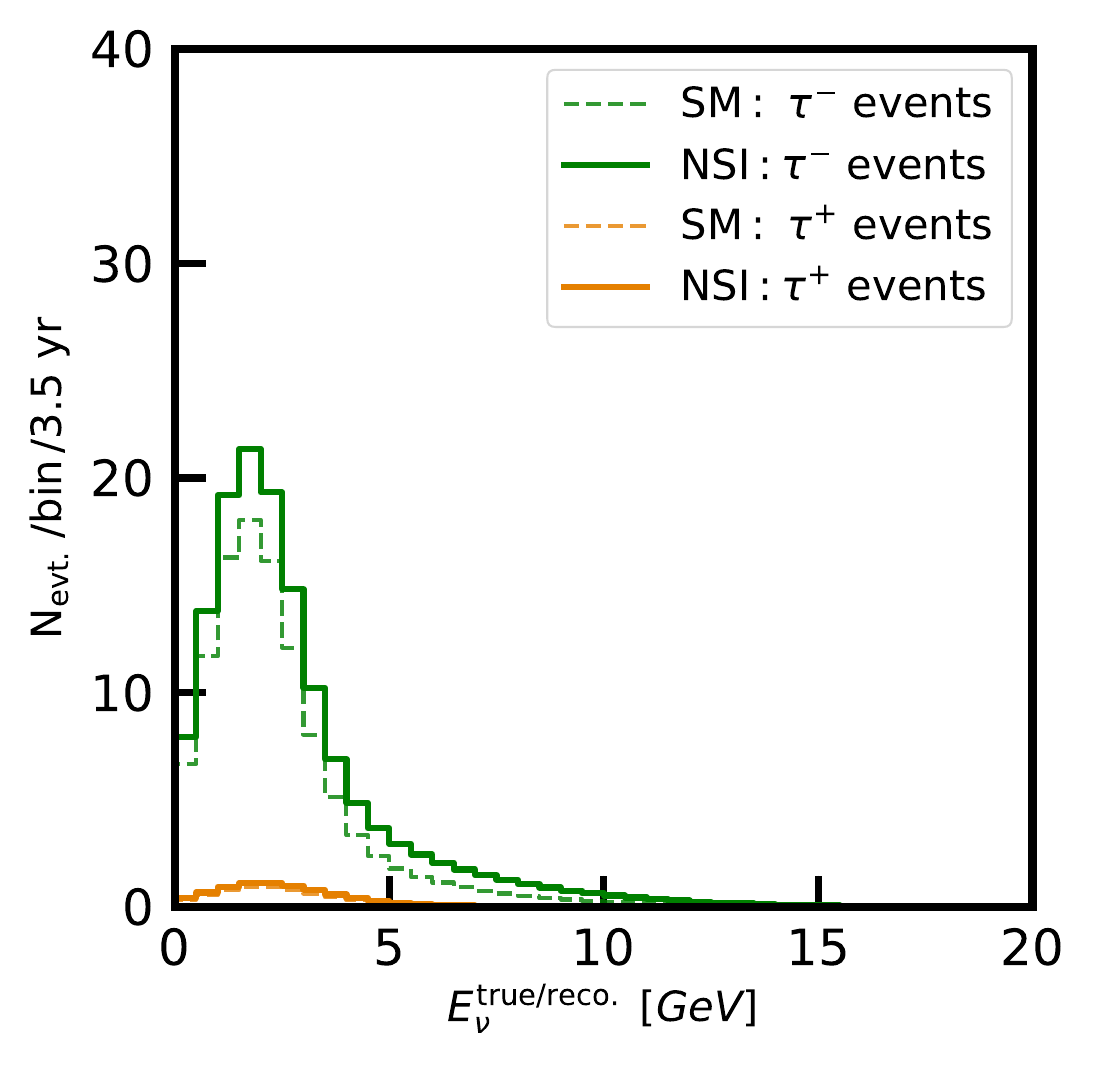}
  \end{minipage}\hfill
  \begin{minipage}{0.32\textwidth}
    \includegraphics[width=\linewidth]{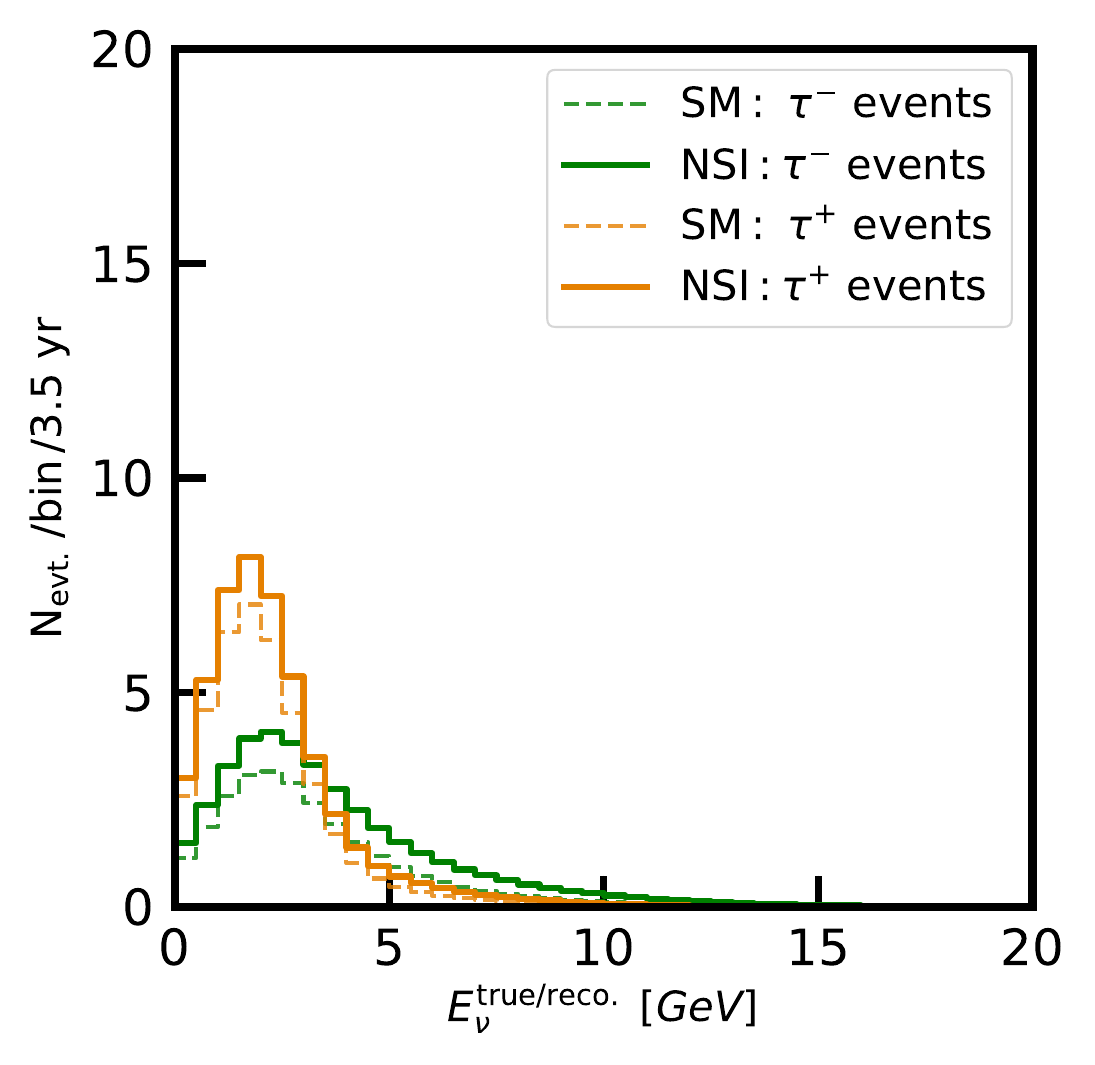}
  \end{minipage}\hfill
  \begin{minipage}{0.32\textwidth}
    \includegraphics[width=\linewidth]{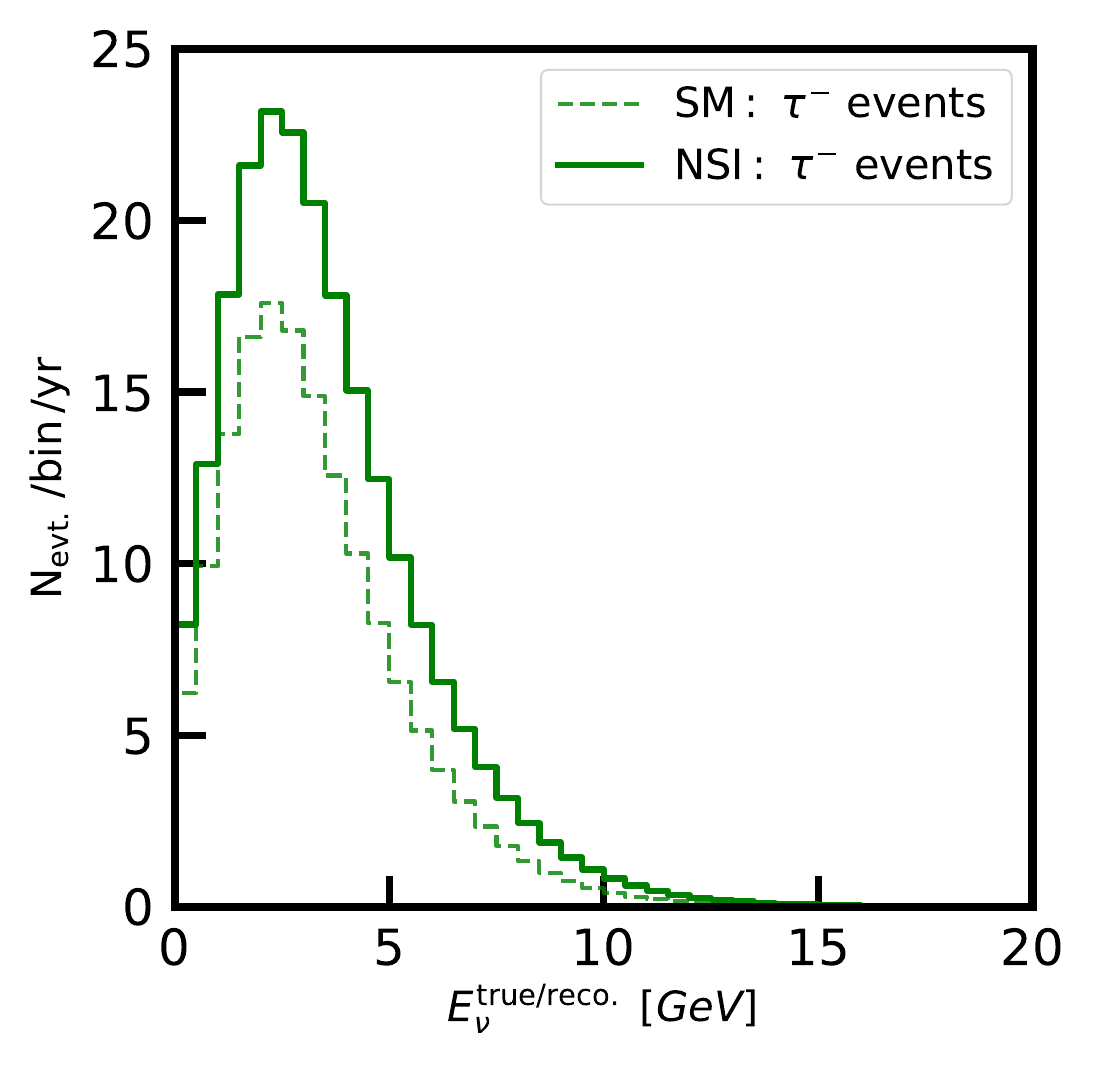}
  \end{minipage}
  \caption{Expected number of events after 3.5 years in neutrino mode (left), 3.5 years in antineutrino mode (middle), 1 year in high energy mode (right). Dashed lines represent the standard scenario ($\epsilon_{\mu\tau}=0$), while solid lines are for $\epsilon_{\mu\tau}=1.2\times10^{-2}$.}
\label{fig:Plot_Event_SM_NSI_reco}
\end{figure}

In a similar way to \cite{DeGouvea:2019kea}, we divided the reconstructed events into 40 energy bins ranging from 0 to 20 GeV, with a constant width of $\Delta E_{\nu} = 0.5$ GeV. The reconstructed SM events at FD, see eq.~(\ref{eq:General_EventSM_reco}), are illustrated in Figure~\ref{fig:Plot_Event_SM_reco} where all the oscillation parameters are fixed at their best-fit values given by eq.~(\ref{eq:NuFit}). The left (middle) graphic shows the $\nu_\tau$($\bar{\nu}_\tau$)-sample to the contributions of $\tau^{\pm}$-reconstruction events for neutrino (antineutrino) mode as solid lines. In dashed lines we also show the true number of events. The right graph illustrates the reconstructed SM events for the high-energy mode as solid lines. As before, dashed lines are reserved for the true events. In order to see the impact that a non-null CC-NSI can have on the expected number of events, we plot in Figure~\ref{fig:Plot_Event_SM_NSI_reco} only the reconstructed events for the three modes when $\epmt=1.2\times10^{-2}$. In this plot, we reserve solid lines for the reconstructed number of events in the presence of CC-NSI, in constract dashed lines show the reconstructed events in the standard scenario ($\epsilon_{\mu\tau}=0$), for comparison. As before, we maintain all oscillation parameters fixed by eq.~(\ref{eq:NuFit}).

\vspace{1cm}
Since CC NSI also modifies the events at the ND, we also performed a study about its sensitivity to $\epmt$. For simplicity, we assumed that the flux can be described using spherical symmetry and that the efficiency in the ND and FD differ only by the difference in their fiducial masses. Thus, we can calculate the events at the ND employing the geometry factor
\begin{eqnarray}
    \kappa
    =
    \frac{
            \phi^{\sm}_{\mu;\:\n}
        }{
            \phi^{\sm}_{\mu;\:\f}
        }
    \frac{
            \sigma_{\tau;\:\n}^{\sm}
        }{
            \sigma_{\tau;\:\f}^{\sm}
        }
    =
    \frac{
            \:M_\n
        }{
            M_\f
        }
    \frac{
            L_\f^2
        }{
            \:L_\n^2
        } ~ ,
\label{eq:Kappa_value}
\end{eqnarray}
where for the DUNE's $X\!=\!\{\n, \f\}$ detector the variable $\phi^{\sm}_{\mu;\:X}$ denotes the $\nu_\mu$ flux at respective detector, $\sigma_{\tau;\:X}^{\sm}$ the cross section for $\nu_\tau$, $M_X$ the fiducial mass, and $L_X^2$ the distance source-detector. Using public data released by the DUNE collaboration~\cite{DUNE:2021cuw,DUNE:2020jqi}, we calculated the ratio of near and far fluxes in order to cross-check the $\kappa/\Delta\sigma_\tau^{\sm}$ value, where we denote $\Delta\sigma_\tau^{\sm} = \sigma_{\tau;\:\n}^{\sm}/\sigma_{\tau;\:\f}^{\sm}$. We checked that the $\kappa/\Delta\sigma_\tau^{\sm}$ ratio is in good agreement with the average ratio of near and far fluxes provided by the collaboration. As a result, the true SM and NSI events at the ND can be estimated from the respective true events at the FD using the relations
\begin{align}
&
    \frac{
            d \NtrSM_{\mt}^{\n}
        }{
            d E_\nu^\tr
        }
    =
    \sbin
    \Big(
    \phi_{\mu;\:\n}^\sm \: 
    \sigma_{\tau;\:\n}^\sm \:
    P_{\mt}^{\n}
    \Big)_i
    =
    \kappa \: 
    \frac{
            P_{\mt}^{\n}
        }{
            \PmtSM
        }
    \:
    \frac{
            d \NtrSM_\mt   
        }{
            d E_\nu^\tr
        } ~ ,
\label{eq:Event_true_Near_SM}
\\
&
    \frac{
            d \NtrNSI_{\mt}^{\n}
        }{
            d E_\nu^\tr
        }
    =
    \sbin
    \lt
    \phi_{\mu;\:\n}^{\sm} \: 
    \sigma_{\tau;\:\n}^{\sm} \:
    \PmtNSI^{\n}
    \rt_i
    =
    \kappa \: 
    \frac{
            \PmtNSI^{\n}
        }{
            \PmtSM
        }
    \:
    \frac{
            d \NtrSM_\mt
        }{
            d E_\nu^\tr
        } ~ ,
\label{eq:Event_true_Near_NSI}
\end{align}
where $\NtrSM_\mt$ are the true SM events at the FD 
eq.~(\ref{eq:EventSM_true}). The upper (lower) relation eq.~(\ref{eq:Event_true_Near_SM}) (eq.~\ref{eq:Event_true_Near_NSI})) generates true SM (NSI) events at the ND. Since the FD and ND detectors are of the same nature, we assumed that the reconstructed events at the ND are produced by the same Gaussian function as the FD. To produce the reconstructed SM and NSI events at the ND, we used the migration matrix eq.~(\ref{eq:Mapping_Matrix}) in the eq.(\ref{eq:Event_true_Near_SM}) and 
eq.~(\ref{eq:Event_true_Near_NSI}), respectively
\begin{align}
&
    \frac{
            d N_{\mt}^{\n}
        }
        {
            d E_\nu^{\reco}
        }
    =
    \int\! dE_{\nu}^{\tr}\: 
    \frac
    { 
        d \NtrSM_{\mt}^{\n}
    }
    {
        dE_{\nu}^{\tr}
    }\:
    f(E_\reco,E_\tr) ~ ,
\label{eq:Event_reco_Near_SM}
\\
&
    \frac{
            d \hat{N}_{\mt}^{\n}
        }
        {
            d E_\nu^{\reco}
        }
    =
    \int\! dE_{\nu}^{\tr}\: 
    \frac
    { 
        d \NtrNSI_{\mt}^{\n}
    }
    {
        dE_{\nu}^{\tr}
    }\:
    f(E_\reco,E_\tr) ~ .
\label{eq:Event_reco_Near_NSI}
\end{align}

Once the reconstructed events are known, we performed an statistic analysis to infer the sensitivity of $\epmt$ from the near and far detectors. The analyses are performed using the $\funchi$ function defined in eq.~(\ref{eq:chi_function}), where the details can be found in the section \ref{sec:numerical}.


\section{Further constraints}
\label{sec:constraints}

In the last section, we have seen how the inclusion of the CC NSI $\epmt$ will modify the event rates of tau neutrinos, to be measured at DUNE. However, a non-null CC NSI will also impact other channels, such as the electron appearance ($\nu_{\mu}\rightarrow \nu_{e}$) and muon disappearance ($\nu_{\mu}\rightarrow \nu_{\mu}$). This can be easily understood by observing eq.~(\ref{eq:rate:final}), where a non-diagonal matrix $\mathcal{P}$ affects arbitrary channels $\nu_{\alpha}\rightarrow\nu_{\beta}$ in general. To probe the impact of a non-null $\epmt$ on the electron appearance and muon disappearance channels, we have used the GLoBES package~\cite{Huber:2004ka,Huber:2007ji}, in conjunction with MonteCUBES~\cite{Blennow:2009pk}. Regarding DUNE simulation, we relied on public GLoBES files supplemented by the DUNE collaboration~\cite{DUNE:2021cuw}, which are based on the Technical Design Report configuration~\cite{DUNE:2020jqi}. For the implementation of CC NSI, we employed internal routines~\cite{Cherchiglia:2023ojf,Cherchiglia:2023aqp}. 
\begin{figure}[!hbt]
    \centering
    \includegraphics[scale=0.3]{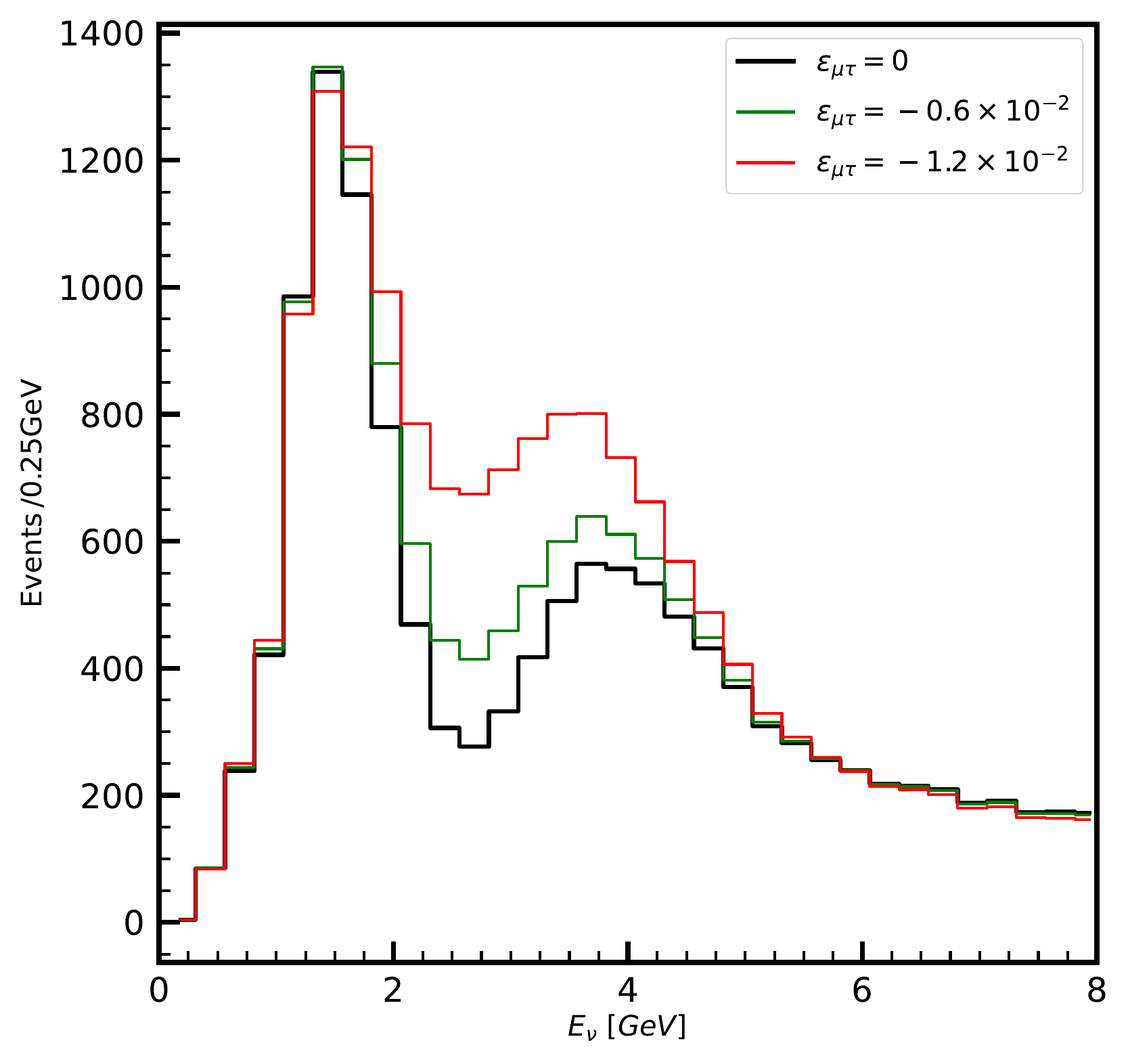}
    \includegraphics[scale=0.3]{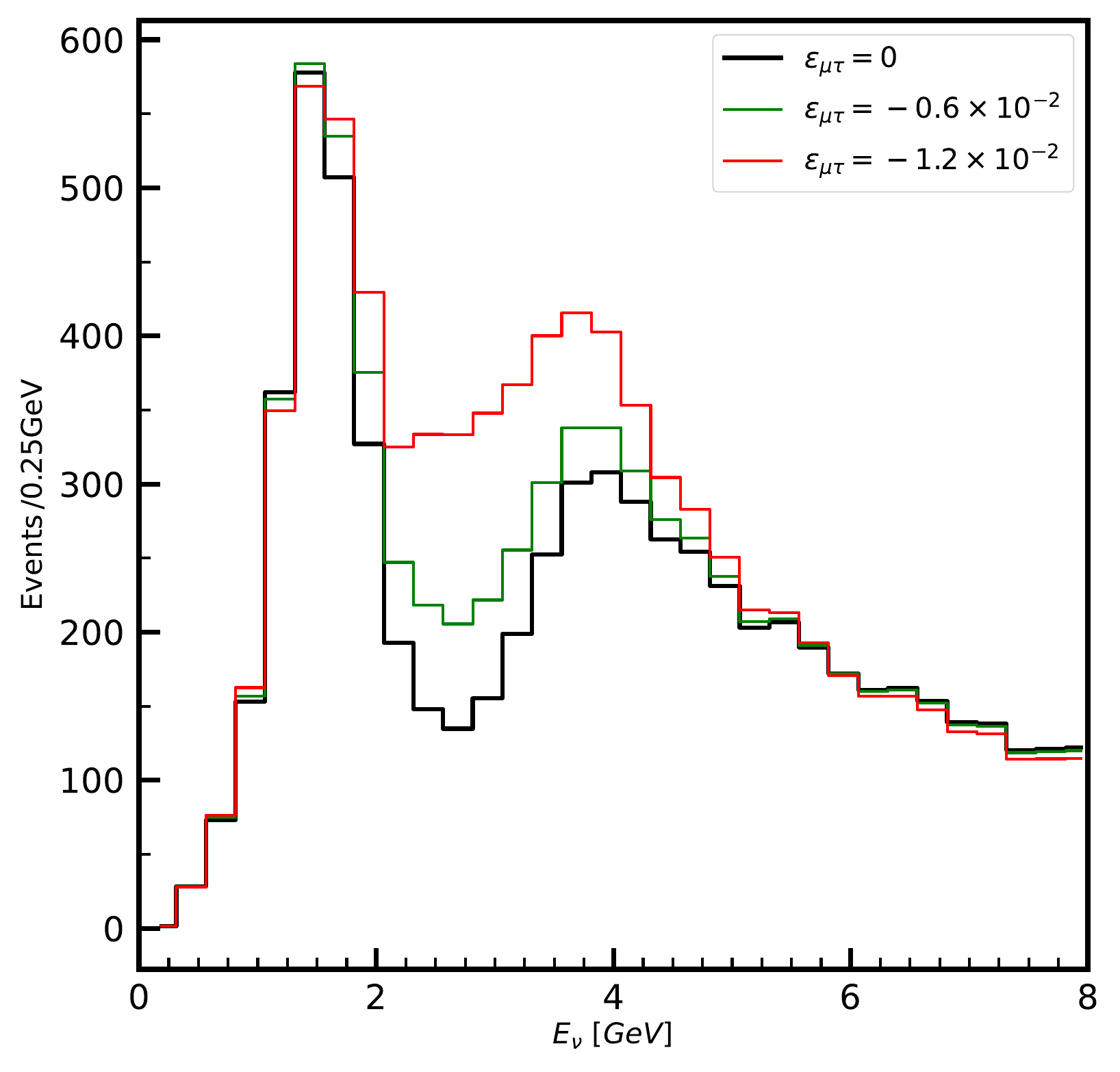}
    \caption{Stacked histograms for neutrino (left) and antineutrino (right) modes. The black curve is for the case of null $\epsilon_{\mu\tau}=0$, while the green (red) curve is for $\epsilon_{\mu\tau}=-6\times 10^{-3} (-1.2 \times 10^{-2})$.}
    \label{fig:events_DUNE}
\end{figure}

For illustration purposes, in Figure \ref{fig:events_DUNE} we show how the number of simulated events is modified in the presence of CC NSI. The left plot is for neutrino mode (5 years) while the right plot is for antineutrino mode (5 years). In black we show the stacking events for the case of null CC NSI, while in green(red) we depict the case with $\epsilon_{\mu\tau}=-6\times 10^{-3} (-1.2 \times 10^{-2})$. In Appendix \ref{ap:prob} we discuss why we choose a negative value for the CC NSI. As can be seen, even for the CC NSI of order $10^{-3}$, a significant increase in the number of events is expected. A thorough statistical analysis is performed in the next section, to obtain the sensitivity on the CC NSI at DUNE, considering only electron appearance and muon disappearance channels. For completeness, we will also consider the sensitivity on $\epsilon_{\mu\tau}$ from the present neutrino experiments, namely T2K and NO$\nu$A, which we present in the next section. For further details of the simulation, we refer the reader to~\cite{Cherchiglia:2023ojf}. 

Finally, the inclusion of CC NSI affects pion decay in general, not only in conjunction with neutrino experiments. The inclusive pion decay into muons can be expressed in our case as \cite{Breso-Pla:2023tnz}
\begin{align}
(\Gamma_{\pi\rightarrow \mu\nu})_{\rm{NSI}}= \frac{V_{ud}^{2}f_{\pi^\pm}^{2}m_\mu^{2}(m_{\pi^{\pm}}^{2}-m_{\mu}^{2})^{2}}{16\pi m_{\pi^{\pm}}^{3}v^{4}}[ {\cal P}{\cal P}^{\dagger}]_{\mu\mu} = \left(\Gamma_{\pi\rightarrow \mu\nu}\right)_{\rm{SM}}(1+p_{\mu}^{2}\epsilon_{\mu\tau}^{2})\,.
\end{align}
Since the decay into electrons is not modified, we can obtain the ratio
\begin{align}
  R_{e/\mu}^{\rm{NSI}}=\frac{R_{e/\mu}^{\rm{theo}}}{(1+p_{\mu}^{2}\epsilon_{\mu\tau}^{2})}  
\end{align}
where $R_{e/\mu}^{\rm{theo}}$ is the theoretical prediction into the SM.

Notice that non-diagonal CC NSI will generally appear in quadratic form, while diagonal CC NSI generate linear terms as well~\cite{Guzzo:2023ayo}. Therefore, the bounds on non-diagonal NSI will be in general weaker. In order to extract the bound on $\epsilon_{\mu\tau}$ we use the ratio $\Gamma_{\pi\rightarrow e\nu}/\Gamma_{\pi\rightarrow \mu\nu}$, whose theoretical~\cite{Cirigliano:2007xi,Bryman:2011zz} and experimental values~\cite{PiENu:2015seu} in the SM are respectively
\begin{align}
R_{e/\mu}^{\rm{theo}}=\left(\frac{\Gamma_{\pi\rightarrow e\nu}}{\Gamma_{\pi\rightarrow \mu\nu}}\right)_{\rm{theo}}=1.2352(2)\times 10^{-4}\,,\quad\quad R_{e/\mu}^{\rm{exp}}=\left(\frac{\Gamma_{\pi\rightarrow e\nu}}{\Gamma_{\pi\rightarrow \mu\nu}}\right)_{\rm{exp}}=1.2344(30)\times 10^{-4}\,.
\end{align}
We obtained the sensitivity on $\epsilon_{\mu\tau}$ by performing a standard chi-squared test
\begin{align}
\chi^{2}=\sum_{\epsilon_{\mu\tau}}\frac{\left(R_{e/\mu}^{\rm{NSI}}-R_{e/\mu}^{\rm{exp}}\right)^{2}}{\sigma^{2}}
\end{align}
where $\sigma$ is the sum in quadrature of the uncertainties in $R_{e/\mu}$.


\section{Numerical Results}
\label{sec:numerical}

As we enter the precision era for neutrino experiments, it may be possible to identify new physics signals from deviations of oscillation parameters in the three-neutrino oscillation model. Naturally, new physics events are expected to be rare and can have non-Gaussian errors. Therefore, we need to consider a method of statistical analysis that converges to the correct value when we have low statistical data and asymmetrical errors. 
Unless the collected events are enormous,  we cannot utilize the usual minimization from the chi-square function. Baker and Cousins defined a $\funchi$ function for fitting functions, adjusting curves to histograms when the bin contents obey a Poisson distribution \cite{Baker:1983tu}. It has a fast convergence, even if the bin contents are very low, and has fewer fluctuations when compared with the chi-square function. The $\funchi$ function can be easily minimized to estimate the parameters and to realize goodness-of-fit testing, behaving asymptotically as the 
classical chi-squared function. Unless otherwise stated, we estimated the parameters through minimization of the Poisson likelihood chi-square defined by~\cite{Almeida:1999ie}
\begin{align}\label{eq:chi_function}
&
    \funchi
    =
    2 \sum_i
    \Big[
        \hat{N}_{i}^{\rm X} - N_{i}^{\rm X}
        +
        N_{i}^{X}
        \log
        \left(
            N_{i}^{\rm X}/\hat{N}_{i}^{\rm X}
        \right) 
    \Big]
    + 
     \left(
               \alpha/\sigma_\alpha
    \right)^2 ~ ,
\\
\nonumber
&
N_{i}^{\rm X}
=
\lt
N_\mt^{\rm X} + N_\text{NC}^X 
\rt_i
~ ,
\quad
\hat{N}_i^{\rm X} 
=
\lt
\hat{N}_{\mt}^X + (1+\alpha)
(1+p_{\mu}^{2}\epmt^2)N_\text{NC}^X 
\rt_i
~ ,
\end{align}
where $N_i^X \ (\hat{N}_i^X)$ is the total number of events without (with) the NSI effect for $i$th-bin 
at $X = \{\n, \f\}$  detector by contribution 
of oscillation signal $\Oscmt$ plus NC 
interaction. We assume that $N_\text{NC}$ in the near detector can be obtained from the far detector simulation by the same proportion defined in eq.~(\ref{eq:Kappa_value}).
For our analysis, the background consists of neutral current (NC) events, which we take from \cite{DeGouvea:2019kea}. For simplicity, we will consider that all systematic uncertainties can be parameterized as a normalization factor for NC events, which we will denote by $(1+\alpha)$. We will treat $\alpha$ as a nuisance parameter, marginalizing it in the range [-1,1]. It is important to notice that, since NC events do not allow for discriminating the neutrino in the final state (no charged lepton is emitted), one has to sum over all flavors. In this case, there is a global factor of $1+p_\mu^2\epsilon_{\mu}^2$, as we show in Appendix \ref{sec:ap}.

\begin{figure}[!hbt]
    \centering
    \includegraphics[scale=0.6]{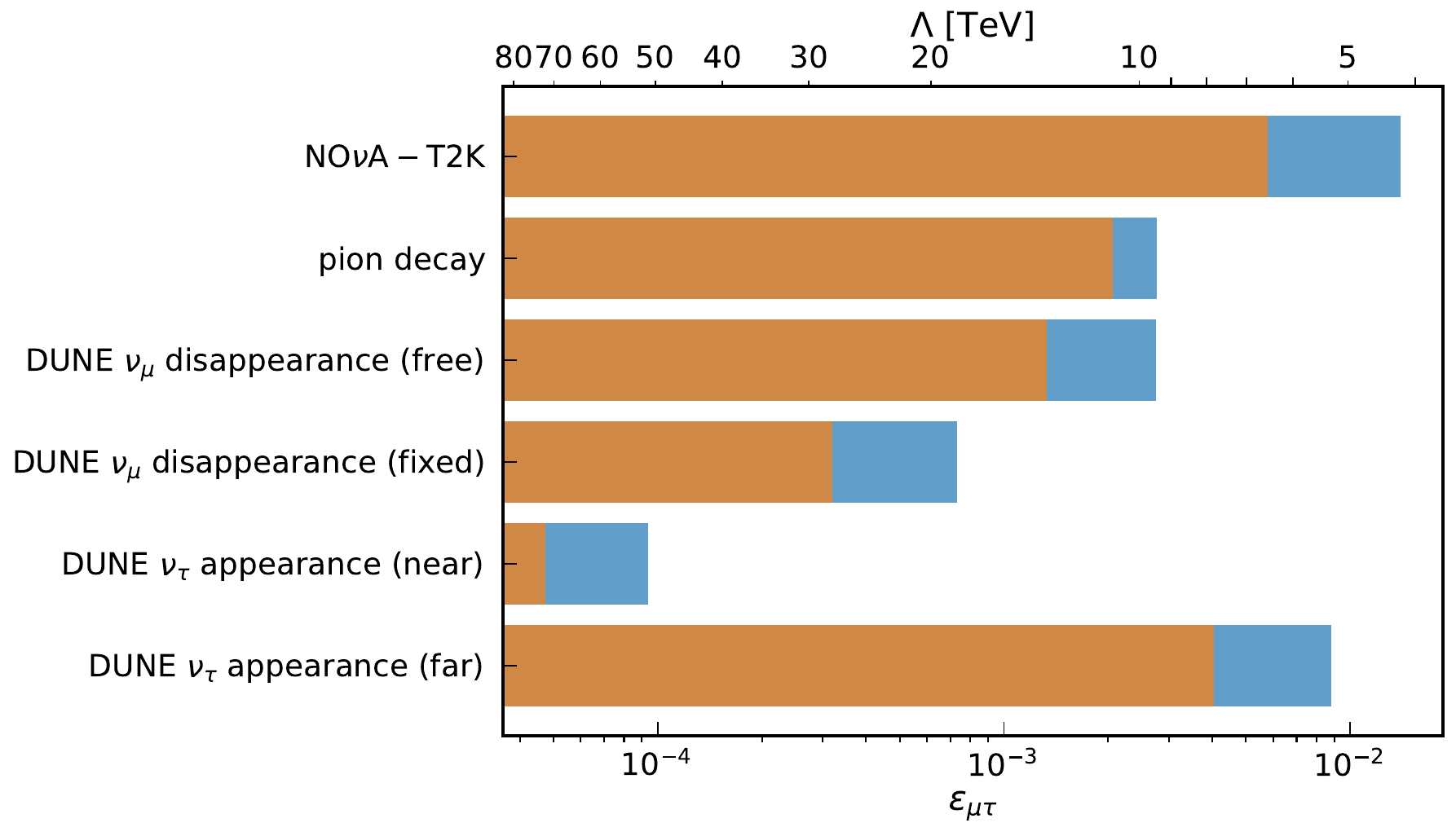}
    \caption{Sensitivity on CC-NSI parameter $\epsilon_{\mu\tau}$ for the different experiments considered in this work. The orange (blue) regions are for one (two) $\sigma$ C.L.}
    \label{fig:chi2_range}
\end{figure}

We show in figure~\ref{fig:chi2_range} the 1$\sigma$ and 2$\sigma$ Confidence Levels (CL) on the modulus of $\epsilon_{\mu\tau}$ for all the experiments described in the last section. For simplicity, in all analyses, we just considered normal 
ordering (NO). As can be seen, the present bounds from neutrino experiments (T2K + NO$\nu$A) are at least one order of magnitude weaker than the present bound coming from pion decay experiments directly. Regarding the electron appearance and muon disappearance channels in DUNE, simulated with GLoBES with a ten-year exposure, we performed two complementary analyses. For the first case, we were interested in the best-case scenario, regarding sensitivity on $\epsilon_{\mu\tau}$. Thus, we have fixed all the oscillation parameters in their best-fit values. This is denoted as fixed scenario. For the second case, we allowed the parameters $\theta_{13}$, $\theta_{23}$, $\delta_{\rm CP}$, $\Delta m_{31}^{2}$ to vary at 1 sigma level. This was denoted by case free scenario. In both cases, we adopted a $5\%$ uncertainty on the average matter density, whose value of $\rho=2.848 \;\unit{g}/\unit{cm}^{3}$ is chosen according to the PREM profile~\cite{Dziewonski:1981xy}. As seen in the figure, a DUNE-like experiment will not, in general, provide more stringent bounds on $\epsilon_{\mu\tau}$ from these channels when compared to the pion decay, unless the precision on the oscillation parameters is improved. 

Finally, for the analysis of tau neutrinos from a DUNE-like experiment (exposure of seven years, one of those at the high-energy run), we provide the sensitivity from measurements of the far and near detectors. Given the smaller number of events, when compared with the electron appearance and muon disappearance channels, the sensitivity on $\epsilon_{\mu\tau}$ is weaker than the case free scenario, for the far detector. Regarding the near detector, however, since in the SM we expect no events, the sensitivity is significantly increased. In particular, a DUNE-like experiment could increase by one order of magnitude the present bound from pion decay.


\section{Concluding Remarks}
\label{sec:conclusion}

In this work, we have investigated the potential of a DUNE-like experiment to probe charged-current non-standard interactions (CC-NSI) through the study of tau neutrino appearance. By considering the impact of CC-NSI in pion decay processes, we demonstrated how modifications in the production mechanism can manifest in the observed event rates at both near and far detectors. Our simulations, based on updated oscillation parameters and reconstructed event spectra, indicate that DUNE has the unique capability to explore regions of parameter space not accessible to current experiments.

While traditional analyses of new physics in neutrino oscillations have emphasized the $\nu_e$ and $\nu_\mu$ channels, our results highlight the complementary role of the $\nu_\tau$ sector. Despite the experimental challenges associated with tau neutrino detection, we have shown that DUNE’s high-energy run can yield competitive sensitivities. In particular, the near detector’s ability to constrain scenarios where the Standard Model predicts no events leads to a projected improvement of more than one order of magnitude over current bounds derived from pion decay. Explicitly, we found the 1~$\sigma$ C.L. limits for the far detector analysis as $\epsilon_{\mu\tau}<4\times10^{-3}$, while for the near detector analysis we found $\epsilon_{\mu\tau}<4.5\times10^{-5}$.

Taken together, these results establish tau neutrino measurements at DUNE as a promising avenue for probing non-standard charged-current interactions. Further refinements in oscillation parameter determinations and dedicated analyses of $\nu_\tau$ samples will be essential to exploit this opportunity fully.

\begin{acknowledgments}

 A.L.C. acknowledges support from the National Council for Scientific and Technological Development – CNPq grant 446121\slash2024-0. O.L.G.P. acknowledges support for FAPESP funding Grant 2014\slash19164-6,2022\slash08954-2, 2021/13757-9 and 2024/07128-7, and the National Council for Scientific and Technological Development – CNPq grant 306565\slash2019-6 and 306405\slash2022-9. A preliminary version of the work was presented in 
 2nd Short-Baseline Experiment-Theory Workshop~\cite{santafe-conference}, Santa Fe, from 2nd to 5th April 2024. This study was financed in part by the Coordenação de Aperfeiçoamento de Pessoal de Nível Superior - Brasil (CAPES) - Finance Code 001.   E. S. S.~acknowledges support from the National Council for Scientific and Technological Development – CNPq through project 140484\slash2023-0.
\end{acknowledgments}


\appendix

\section{Influence of CC-NSI to the NC background}
\label{sec:ap}

For our work, the background is given by NC interactions with the medium. In this case, there is a coherent sum of all types of neutrinos in the final state. In the presence of CC-NSI, the same reasoning still applies. Thus, we need to sum the probabilities for $\alpha = e,\mu,\tau$,

\begin{align}
\sum_{\alpha=e,\mu,\tau} P_{\mu\to\alpha}
&=
\sum_{\alpha}
\left|
S_{\alpha \mu}^{\rm SM}
-
p_\mu \, \epsilon_{\mu\tau}^{*} \,
S_{\alpha \tau}^{\rm SM}
\right|^2.
\label{eq:sumP_def}
\end{align}

Expanding the modulus squared, we obtain
\begin{align}
\sum_{\alpha} P_{\mu\to\alpha}
&=
\sum_{\alpha}
\Big(
|S_{\alpha\mu}^{\rm SM}|^2
+
|p_\mu|^2 |\epsilon_{\mu\tau}|^2 |S_{\alpha\tau}^{\rm SM}|^2
-
p_\mu \epsilon_{\mu\tau}^{*} S_{\alpha\mu}^{\rm SM*} S_{\alpha\tau}^{\rm SM}
-
p_\mu^{*} \epsilon_{\mu\tau} S_{\alpha\mu}^{\rm SM} S_{\alpha\tau}^{\rm SM*}
\Big).
\label{eq:sumP_expand}
\end{align}

Using the unitarity of the evolution matrix  in the SM,
\begin{align}
\sum_{\alpha} |S_{\alpha\mu}^{\rm SM}|^2 &= 1,
&
\sum_{\alpha} |S_{\alpha\tau}^{\rm SM}|^2 &= 1,
&
\sum_{\alpha} S_{\alpha\mu}^{\rm SM} S_{\alpha\tau}^{\rm SM*} &= \sum_{\alpha} \left(S^{\dagger}\right)^{\rm SM}_{\tau\alpha} S_{\alpha\mu}^{\rm SM}= 0
\quad (\mu \neq \tau),
\end{align}
the interference terms cancel, yielding
\begin{align}
\sum_{\alpha=e,\mu,\tau} P_{\mu\to\alpha}
=
1
+
|p_\mu|^2 \, |\epsilon_{\mu\tau}|^2.
\label{eq:sumP_result}
\end{align}

\section{Probability for the $\nu_{\mu}$ disappearance channel}
\label{ap:prob}

We briefly provide further details regarding the muon neutrino disappearance channel at DUNE. In this case the disappearance probability in the presence of NSI is given by 

\begin{align}
P_{\mu\mu}^{\nsi}
    =
    \left|
        S_{\mu \mu}^{\sm} 
        - 
        p_\mu 
        \epmt^{*}
        S_{\mu \tau}^{\sm}
    \right|^2 ~ .
\end{align}

  \begin{figure}[h!]
        \centering
        \includegraphics[scale=0.4]{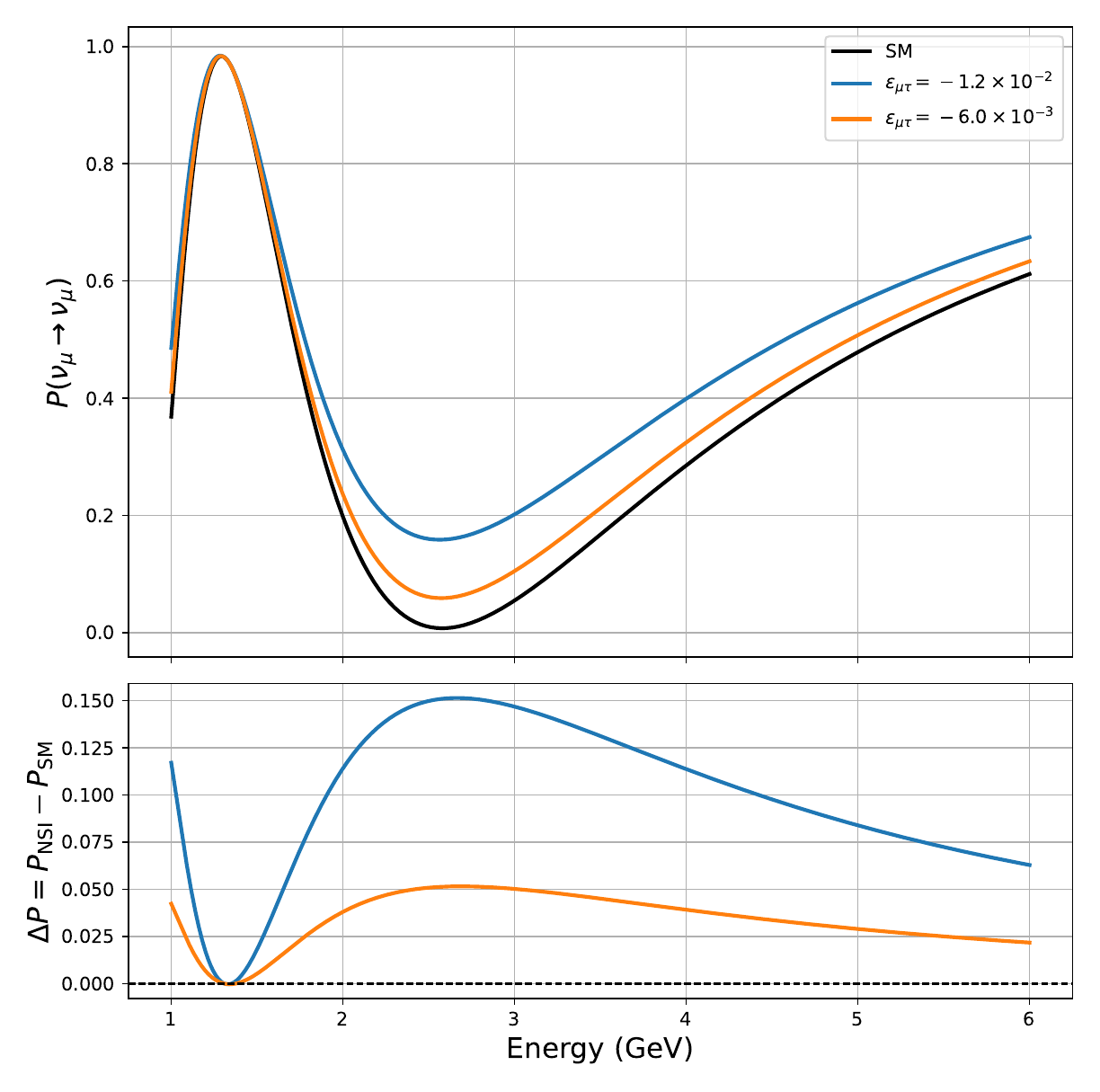}
        \includegraphics[scale=0.4]{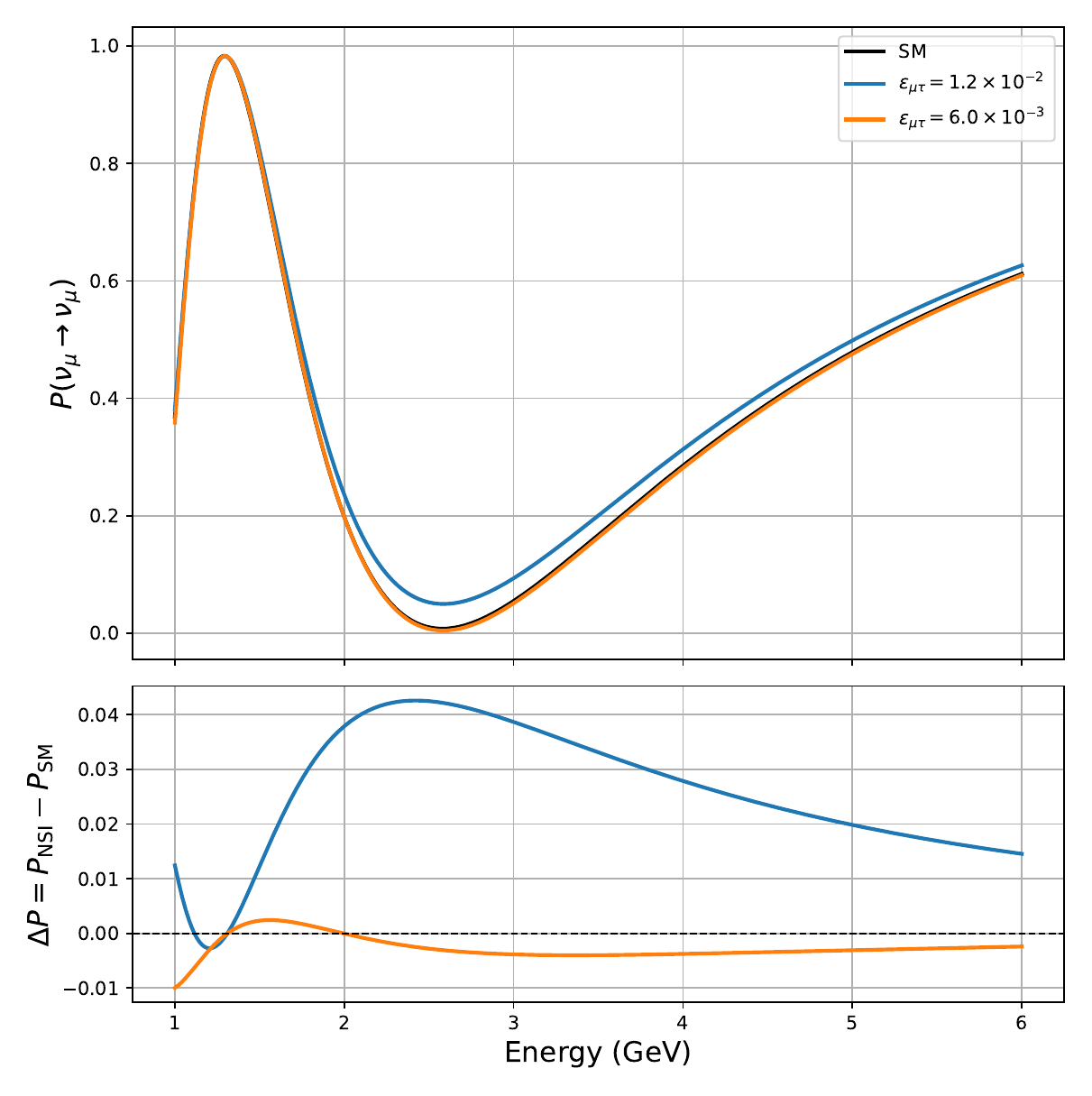}
        \caption{Probabilities for tau appearance for positive and negative values of $\epmt$.}
        \label{fig:prob}
    \end{figure}

Keeping all standard oscillation parameters fixed in their best-fit values, we show in figure \ref{fig:prob} the tau neutrino appearance probability for the SM (black line), and for $\epmt=0.6(1.2)\times10^{-2}$ as orange(blue) lines. We also show the difference between the NSI and the SM cases. It is interesting to notice that for $\epmt=\pm1.2\times10^{-2}$, the NSI probability is larger than the SM one for almost all energies values, the only exemption being around an energy of 1.29 GeV. This happens because $P_{\mu\mu}^{\sm}=\left|S_{\mu \mu}^{\sm}
    \right|^2$ is maximum for this energy, which implies that $\left|S_{\mu \tau}^{\sm}
    \right|^2$ is very suppressed. Therefore, for this energy, the probability in the presence of NSI is approximately given by
\begin{align}
P_{\mu\mu}^{\nsi}
    \sim
    \left|
        S_{\mu \mu}^{\sm}\right|^2 
        -2 
        p_\mu 
        \epmt \Re\left[S_{\mu \mu}^{\sm}
        S_{\mu \tau}^{\sm*}\right]
    ~ .
\end{align}

In other words, the linear term in $\epmt$ will be the leading one for this energy, explaining the difference in behavior around 1.29 GeV among the curves with positive and negative values for $\epmt=1.2\times10^{-2}$. For $\epmt=\pm0.6\times10^{-2}$, however, there is a very different behavior for all energies values, since in this case, the linear term is more relevant. 

\begin{figure}[!h]
    \centering
    \includegraphics[scale=0.3]{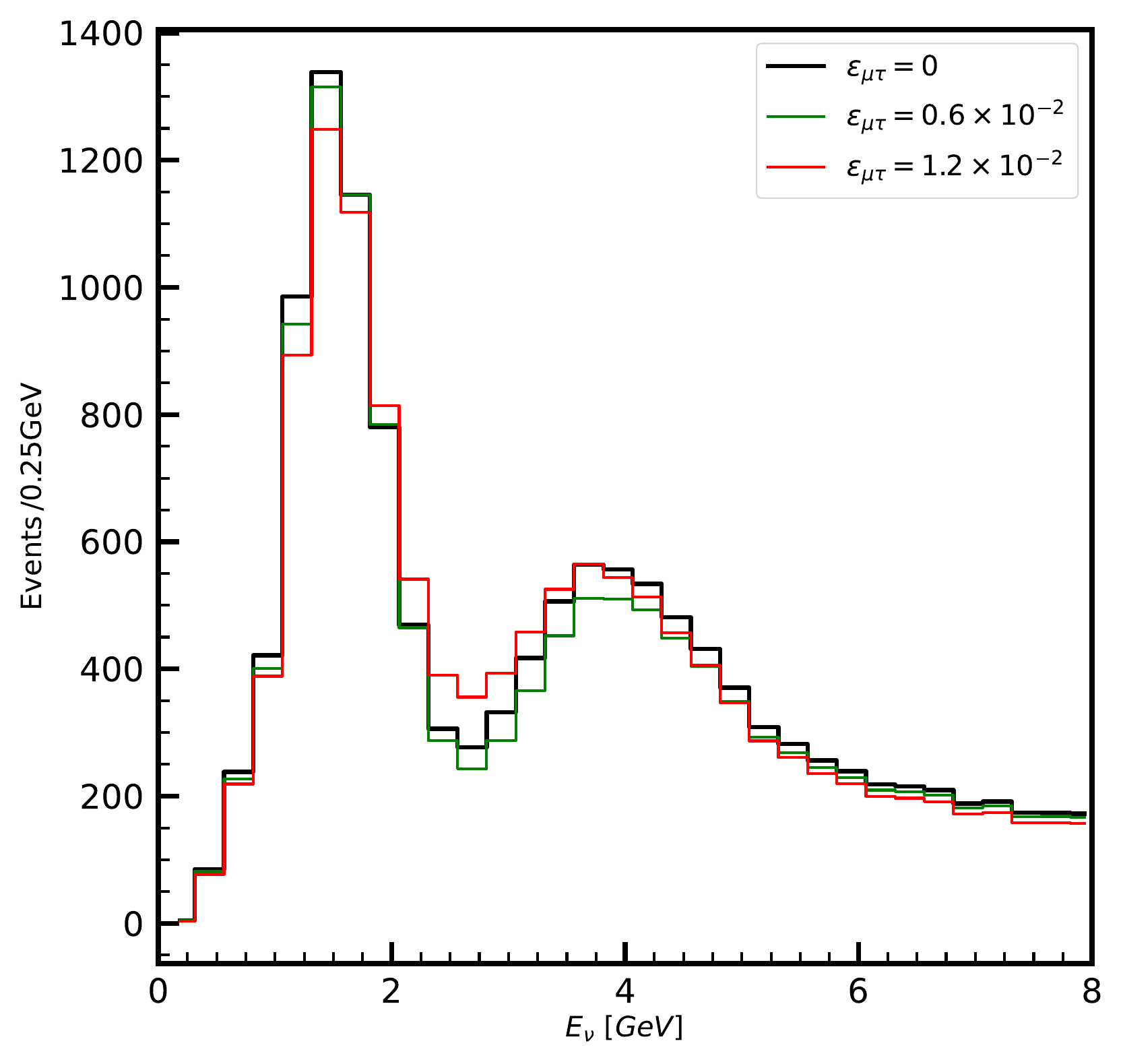}
    \caption{Stacked histograms for neutrino (left) and antineutrino (right) modes. The black curve is for the case of null $\epsilon_{\mu\tau}=0$, while the green (red) curve is for $\epsilon_{\mu\tau}=6\times 10^{-3} (1.2 \times 10^{-2})$.}
    \label{fig:events_DUNE2}
\end{figure}

Another interesting conclusion that we can draw from these plots is that negative values of $\epmt$ give potentially larger signals. For comparison, if we have chosen an positive $\epmt$ in section \ref{sec:constraints}, the left plot of figure \ref{fig:events_DUNE} would be replaced by figure \ref{fig:events_DUNE2}

\bibliographystyle{apsrev4-1}
\bibliography{draft}

\begin{thebibliography}{68}%
\makeatletter
\providecommand \@ifxundefined [1]{%
 \@ifx{#1\undefined}
}%
\providecommand \@ifnum [1]{%
 \ifnum #1\expandafter \@firstoftwo
 \else \expandafter \@secondoftwo
 \fi
}%
\providecommand \@ifx [1]{%
 \ifx #1\expandafter \@firstoftwo
 \else \expandafter \@secondoftwo
 \fi
}%
\providecommand \natexlab [1]{#1}%
\providecommand \enquote  [1]{``#1''}%
\providecommand \bibnamefont  [1]{#1}%
\providecommand \bibfnamefont [1]{#1}%
\providecommand \citenamefont [1]{#1}%
\providecommand \href@noop [0]{\@secondoftwo}%
\providecommand \href [0]{\begingroup \@sanitize@url \@href}%
\providecommand \@href[1]{\@@startlink{#1}\@@href}%
\providecommand \@@href[1]{\endgroup#1\@@endlink}%
\providecommand \@sanitize@url [0]{\catcode `\\12\catcode `\$12\catcode
  `\&12\catcode `\#12\catcode `\^12\catcode `\_12\catcode `\%12\relax}%
\providecommand \@@startlink[1]{}%
\providecommand \@@endlink[0]{}%
\providecommand \url  [0]{\begingroup\@sanitize@url \@url }%
\providecommand \@url [1]{\endgroup\@href {#1}{\urlprefix }}%
\providecommand \urlprefix  [0]{URL }%
\providecommand \Eprint [0]{\href }%
\providecommand \doibase [0]{http://dx.doi.org/}%
\providecommand \selectlanguage [0]{\@gobble}%
\providecommand \bibinfo  [0]{\@secondoftwo}%
\providecommand \bibfield  [0]{\@secondoftwo}%
\providecommand \translation [1]{[#1]}%
\providecommand \BibitemOpen [0]{}%
\providecommand \bibitemStop [0]{}%
\providecommand \bibitemNoStop [0]{.\EOS\space}%
\providecommand \EOS [0]{\spacefactor3000\relax}%
\providecommand \BibitemShut  [1]{\csname bibitem#1\endcsname}%
\let\auto@bib@innerbib\@empty
\bibitem [{\citenamefont {Abi}\ \emph {et~al.}(2020{\natexlab{a}})\citenamefont
  {Abi} \emph {et~al.}}]{DUNE:2020ypp}%
  \BibitemOpen
  \bibfield  {author} {\bibinfo {author} {\bibfnamefont {B.}~\bibnamefont
  {Abi}} \emph {et~al.} (\bibinfo {collaboration} {DUNE}),\ }\href@noop {} {\
  (\bibinfo {year} {2020}{\natexlab{a}})},\ \Eprint
  {http://arxiv.org/abs/2002.03005} {arXiv:2002.03005 [hep-ex]} \BibitemShut
  {NoStop}%
\bibitem [{\citenamefont {Abed~Abud}\ \emph {et~al.}(2024)\citenamefont
  {Abed~Abud} \emph {et~al.}}]{DUNE:2024wvj}%
  \BibitemOpen
  \bibfield  {author} {\bibinfo {author} {\bibfnamefont {A.}~\bibnamefont
  {Abed~Abud}} \emph {et~al.} (\bibinfo {collaboration} {DUNE}),\ }\href
  {\doibase 10.1088/1748-0221/19/12/P12005} {\bibfield  {journal} {\bibinfo
  {journal} {JINST}\ }\textbf {\bibinfo {volume} {19}},\ \bibinfo {pages}
  {P12005} (\bibinfo {year} {2024})},\ \Eprint
  {http://arxiv.org/abs/2408.12725} {arXiv:2408.12725 [physics.ins-det]}
  \BibitemShut {NoStop}%
\bibitem [{\citenamefont {Arg\"uelles}\ \emph {et~al.}(2023)\citenamefont
  {Arg\"uelles} \emph {et~al.}}]{Arguelles:2022tki}%
  \BibitemOpen
  \bibfield  {author} {\bibinfo {author} {\bibfnamefont {C.~A.}\ \bibnamefont
  {Arg\"uelles}} \emph {et~al.},\ }\href {\doibase
  10.1140/epjc/s10052-022-11049-7} {\bibfield  {journal} {\bibinfo  {journal}
  {Eur. Phys. J. C}\ }\textbf {\bibinfo {volume} {83}},\ \bibinfo {pages} {15}
  (\bibinfo {year} {2023})},\ \Eprint {http://arxiv.org/abs/2203.10811}
  {arXiv:2203.10811 [hep-ph]} \BibitemShut {NoStop}%
\bibitem [{\citenamefont {Giunti}(2002)}]{Giunti:2001ws}%
  \BibitemOpen
  \bibfield  {author} {\bibinfo {author} {\bibfnamefont {C.}~\bibnamefont
  {Giunti}},\ }\href {\doibase 10.1103/PhysRevD.65.033006} {\bibfield
  {journal} {\bibinfo  {journal} {Phys. Rev. D}\ }\textbf {\bibinfo {volume}
  {65}},\ \bibinfo {pages} {033006} (\bibinfo {year} {2002})},\ \Eprint
  {http://arxiv.org/abs/hep-ph/0107310} {arXiv:hep-ph/0107310} \BibitemShut
  {NoStop}%
\bibitem [{\citenamefont {Bertou}\ \emph {et~al.}(2002)\citenamefont {Bertou},
  \citenamefont {Billoir}, \citenamefont {Deligny}, \citenamefont {Lachaud},\
  and\ \citenamefont {Letessier-Selvon}}]{Bertou:2001vm}%
  \BibitemOpen
  \bibfield  {author} {\bibinfo {author} {\bibfnamefont {X.}~\bibnamefont
  {Bertou}}, \bibinfo {author} {\bibfnamefont {P.}~\bibnamefont {Billoir}},
  \bibinfo {author} {\bibfnamefont {O.}~\bibnamefont {Deligny}}, \bibinfo
  {author} {\bibfnamefont {C.}~\bibnamefont {Lachaud}}, \ and\ \bibinfo
  {author} {\bibfnamefont {A.}~\bibnamefont {Letessier-Selvon}},\ }\href
  {\doibase 10.1016/S0927-6505(01)00147-5} {\bibfield  {journal} {\bibinfo
  {journal} {Astropart. Phys.}\ }\textbf {\bibinfo {volume} {17}},\ \bibinfo
  {pages} {183} (\bibinfo {year} {2002})},\ \Eprint
  {http://arxiv.org/abs/astro-ph/0104452} {arXiv:astro-ph/0104452} \BibitemShut
  {NoStop}%
\bibitem [{\citenamefont {De~Gouv\^ea}\ \emph {et~al.}(2019)\citenamefont
  {De~Gouv\^ea}, \citenamefont {Kelly}, \citenamefont {Stenico},\ and\
  \citenamefont {Pasquini}}]{DeGouvea:2019kea}%
  \BibitemOpen
  \bibfield  {author} {\bibinfo {author} {\bibfnamefont {A.}~\bibnamefont
  {De~Gouv\^ea}}, \bibinfo {author} {\bibfnamefont {K.~J.}\ \bibnamefont
  {Kelly}}, \bibinfo {author} {\bibfnamefont {G.~V.}\ \bibnamefont {Stenico}},
  \ and\ \bibinfo {author} {\bibfnamefont {P.}~\bibnamefont {Pasquini}},\
  }\href {\doibase 10.1103/PhysRevD.100.016004} {\bibfield  {journal} {\bibinfo
   {journal} {Phys. Rev. D}\ }\textbf {\bibinfo {volume} {100}},\ \bibinfo
  {pages} {016004} (\bibinfo {year} {2019})},\ \Eprint
  {http://arxiv.org/abs/1904.07265} {arXiv:1904.07265 [hep-ph]} \BibitemShut
  {NoStop}%
\bibitem [{\citenamefont {Farzan}(2021)}]{Farzan:2021gbx}%
  \BibitemOpen
  \bibfield  {author} {\bibinfo {author} {\bibfnamefont {Y.}~\bibnamefont
  {Farzan}},\ }\href {\doibase 10.1007/JHEP07(2021)174} {\bibfield  {journal}
  {\bibinfo  {journal} {JHEP}\ }\textbf {\bibinfo {volume} {07}},\ \bibinfo
  {pages} {174} (\bibinfo {year} {2021})},\ \Eprint
  {http://arxiv.org/abs/2105.03272} {arXiv:2105.03272 [hep-ph]} \BibitemShut
  {NoStop}%
\bibitem [{\citenamefont {Saveliev}\ and\ \citenamefont
  {Hyde}(2022)}]{Saveliev:2021jtw}%
  \BibitemOpen
  \bibfield  {author} {\bibinfo {author} {\bibfnamefont {M.}~\bibnamefont
  {Saveliev}}\ and\ \bibinfo {author} {\bibfnamefont {J.}~\bibnamefont
  {Hyde}},\ }\href {\doibase 10.1103/PhysRevD.105.083025} {\bibfield  {journal}
  {\bibinfo  {journal} {Phys. Rev. D}\ }\textbf {\bibinfo {volume} {105}},\
  \bibinfo {pages} {083025} (\bibinfo {year} {2022})},\ \Eprint
  {http://arxiv.org/abs/2108.13412} {arXiv:2108.13412 [hep-ph]} \BibitemShut
  {NoStop}%
\bibitem [{\citenamefont {Martinez-Soler}\ and\ \citenamefont
  {Minakata}(2021)}]{Martinez-Soler:2021sir}%
  \BibitemOpen
  \bibfield  {author} {\bibinfo {author} {\bibfnamefont {I.}~\bibnamefont
  {Martinez-Soler}}\ and\ \bibinfo {author} {\bibfnamefont {H.}~\bibnamefont
  {Minakata}},\ }\href {\doibase 10.1103/PhysRevD.104.093006} {\bibfield
  {journal} {\bibinfo  {journal} {Phys. Rev. D}\ }\textbf {\bibinfo {volume}
  {104}},\ \bibinfo {pages} {093006} (\bibinfo {year} {2021})},\ \Eprint
  {http://arxiv.org/abs/2109.06933} {arXiv:2109.06933 [hep-ph]} \BibitemShut
  {NoStop}%
\bibitem [{\citenamefont {Denton}(2021)}]{Denton:2021rsa}%
  \BibitemOpen
  \bibfield  {author} {\bibinfo {author} {\bibfnamefont {P.~B.}\ \bibnamefont
  {Denton}},\ }\href {\doibase 10.1103/PhysRevD.104.113003} {\bibfield
  {journal} {\bibinfo  {journal} {Phys. Rev. D}\ }\textbf {\bibinfo {volume}
  {104}},\ \bibinfo {pages} {113003} (\bibinfo {year} {2021})},\ \Eprint
  {http://arxiv.org/abs/2109.14576} {arXiv:2109.14576 [hep-ph]} \BibitemShut
  {NoStop}%
\bibitem [{\citenamefont {Ansarifard}\ and\ \citenamefont
  {Farzan}(2022)}]{Ansarifard:2021dju}%
  \BibitemOpen
  \bibfield  {author} {\bibinfo {author} {\bibfnamefont {S.}~\bibnamefont
  {Ansarifard}}\ and\ \bibinfo {author} {\bibfnamefont {Y.}~\bibnamefont
  {Farzan}},\ }\href {\doibase 10.1140/epjc/s10052-022-10512-9} {\bibfield
  {journal} {\bibinfo  {journal} {Eur. Phys. J. C}\ }\textbf {\bibinfo {volume}
  {82}},\ \bibinfo {pages} {568} (\bibinfo {year} {2022})},\ \Eprint
  {http://arxiv.org/abs/2112.08799} {arXiv:2112.08799 [hep-ph]} \BibitemShut
  {NoStop}%
\bibitem [{\citenamefont {Huang}\ \emph {et~al.}(2022)\citenamefont {Huang},
  \citenamefont {Jana}, \citenamefont {Lindner},\ and\ \citenamefont
  {Rodejohann}}]{Huang:2021mki}%
  \BibitemOpen
  \bibfield  {author} {\bibinfo {author} {\bibfnamefont {G.-y.}\ \bibnamefont
  {Huang}}, \bibinfo {author} {\bibfnamefont {S.}~\bibnamefont {Jana}},
  \bibinfo {author} {\bibfnamefont {M.}~\bibnamefont {Lindner}}, \ and\
  \bibinfo {author} {\bibfnamefont {W.}~\bibnamefont {Rodejohann}},\ }\href
  {\doibase 10.1088/1475-7516/2022/02/038} {\bibfield  {journal} {\bibinfo
  {journal} {JCAP}\ }\textbf {\bibinfo {volume} {02}},\ \bibinfo {pages} {038}
  (\bibinfo {year} {2022})},\ \Eprint {http://arxiv.org/abs/2112.09476}
  {arXiv:2112.09476 [hep-ph]} \BibitemShut {NoStop}%
\bibitem [{\citenamefont {Mammen~Abraham}\ \emph {et~al.}(2022)\citenamefont
  {Mammen~Abraham} \emph {et~al.}}]{MammenAbraham:2022xoc}%
  \BibitemOpen
  \bibfield  {author} {\bibinfo {author} {\bibfnamefont {R.}~\bibnamefont
  {Mammen~Abraham}} \emph {et~al.},\ }\href {\doibase 10.1088/1361-6471/ac89d2}
  {\bibfield  {journal} {\bibinfo  {journal} {J. Phys. G}\ }\textbf {\bibinfo
  {volume} {49}},\ \bibinfo {pages} {110501} (\bibinfo {year} {2022})},\
  \Eprint {http://arxiv.org/abs/2203.05591} {arXiv:2203.05591 [hep-ph]}
  \BibitemShut {NoStop}%
\bibitem [{\citenamefont {Huang}(2022)}]{Huang:2022ebg}%
  \BibitemOpen
  \bibfield  {author} {\bibinfo {author} {\bibfnamefont {G.-y.}\ \bibnamefont
  {Huang}},\ }\href {\doibase 10.1140/epjc/s10052-022-11052-y} {\bibfield
  {journal} {\bibinfo  {journal} {Eur. Phys. J. C}\ }\textbf {\bibinfo {volume}
  {82}},\ \bibinfo {pages} {1089} (\bibinfo {year} {2022})},\ \Eprint
  {http://arxiv.org/abs/2207.02222} {arXiv:2207.02222 [hep-ph]} \BibitemShut
  {NoStop}%
\bibitem [{\citenamefont {Bakhti}\ \emph {et~al.}(2024)\citenamefont {Bakhti},
  \citenamefont {Rajaee},\ and\ \citenamefont {Shin}}]{Bakhti:2023mvo}%
  \BibitemOpen
  \bibfield  {author} {\bibinfo {author} {\bibfnamefont {P.}~\bibnamefont
  {Bakhti}}, \bibinfo {author} {\bibfnamefont {M.}~\bibnamefont {Rajaee}}, \
  and\ \bibinfo {author} {\bibfnamefont {S.}~\bibnamefont {Shin}},\ }\href
  {\doibase 10.1103/PhysRevD.109.095043} {\bibfield  {journal} {\bibinfo
  {journal} {Phys. Rev. D}\ }\textbf {\bibinfo {volume} {109}},\ \bibinfo
  {pages} {095043} (\bibinfo {year} {2024})},\ \Eprint
  {http://arxiv.org/abs/2311.14945} {arXiv:2311.14945 [hep-ph]} \BibitemShut
  {NoStop}%
\bibitem [{\citenamefont {Meighen-Berger}\ \emph {et~al.}(2024)\citenamefont
  {Meighen-Berger}, \citenamefont {Beacom}, \citenamefont {Bell},\ and\
  \citenamefont {Dolan}}]{Meighen-Berger:2023xpr}%
  \BibitemOpen
  \bibfield  {author} {\bibinfo {author} {\bibfnamefont {S.~A.}\ \bibnamefont
  {Meighen-Berger}}, \bibinfo {author} {\bibfnamefont {J.~F.}\ \bibnamefont
  {Beacom}}, \bibinfo {author} {\bibfnamefont {N.~F.}\ \bibnamefont {Bell}}, \
  and\ \bibinfo {author} {\bibfnamefont {M.~J.}\ \bibnamefont {Dolan}},\ }\href
  {\doibase 10.1103/PhysRevD.109.092006} {\bibfield  {journal} {\bibinfo
  {journal} {Phys. Rev. D}\ }\textbf {\bibinfo {volume} {109}},\ \bibinfo
  {pages} {092006} (\bibinfo {year} {2024})},\ \Eprint
  {http://arxiv.org/abs/2311.01667} {arXiv:2311.01667 [hep-ph]} \BibitemShut
  {NoStop}%
\bibitem [{\citenamefont {Dev}\ \emph {et~al.}(2024)\citenamefont {Dev},
  \citenamefont {Dutta}, \citenamefont {Han},\ and\ \citenamefont
  {Kim}}]{Dev:2023rqb}%
  \BibitemOpen
  \bibfield  {author} {\bibinfo {author} {\bibfnamefont {P.~S.~B.}\
  \bibnamefont {Dev}}, \bibinfo {author} {\bibfnamefont {B.}~\bibnamefont
  {Dutta}}, \bibinfo {author} {\bibfnamefont {T.}~\bibnamefont {Han}}, \ and\
  \bibinfo {author} {\bibfnamefont {D.}~\bibnamefont {Kim}},\ }\href {\doibase
  10.1016/j.physletb.2024.138500} {\bibfield  {journal} {\bibinfo  {journal}
  {Phys. Lett. B}\ }\textbf {\bibinfo {volume} {850}},\ \bibinfo {pages}
  {138500} (\bibinfo {year} {2024})},\ \Eprint
  {http://arxiv.org/abs/2304.02031} {arXiv:2304.02031 [hep-ph]} \BibitemShut
  {NoStop}%
\bibitem [{\citenamefont {Abbasi}\ \emph {et~al.}(2024)\citenamefont {Abbasi}
  \emph {et~al.}}]{IceCube:2024nhk}%
  \BibitemOpen
  \bibfield  {author} {\bibinfo {author} {\bibfnamefont {R.}~\bibnamefont
  {Abbasi}} \emph {et~al.} (\bibinfo {collaboration} {IceCube}),\ }\href
  {\doibase 10.1103/PhysRevLett.132.151001} {\bibfield  {journal} {\bibinfo
  {journal} {Phys. Rev. Lett.}\ }\textbf {\bibinfo {volume} {132}},\ \bibinfo
  {pages} {151001} (\bibinfo {year} {2024})},\ \Eprint
  {http://arxiv.org/abs/2403.02516} {arXiv:2403.02516 [astro-ph.HE]}
  \BibitemShut {NoStop}%
\bibitem [{\citenamefont {Francener}\ \emph
  {et~al.}(2024{\natexlab{a}})\citenamefont {Francener}, \citenamefont
  {Goncalves},\ and\ \citenamefont {Gratieri}}]{Francener:2024ney}%
  \BibitemOpen
  \bibfield  {author} {\bibinfo {author} {\bibfnamefont {R.}~\bibnamefont
  {Francener}}, \bibinfo {author} {\bibfnamefont {V.~P.}\ \bibnamefont
  {Goncalves}}, \ and\ \bibinfo {author} {\bibfnamefont {D.~R.}\ \bibnamefont
  {Gratieri}},\ }\href {\doibase 10.1103/PhysRevD.109.113005} {\bibfield
  {journal} {\bibinfo  {journal} {Phys. Rev. D}\ }\textbf {\bibinfo {volume}
  {109}},\ \bibinfo {pages} {113005} (\bibinfo {year} {2024}{\natexlab{a}})},\
  \Eprint {http://arxiv.org/abs/2405.08508} {arXiv:2405.08508 [hep-ph]}
  \BibitemShut {NoStop}%
\bibitem [{\citenamefont {Francener}\ \emph
  {et~al.}(2024{\natexlab{b}})\citenamefont {Francener}, \citenamefont
  {Goncalves},\ and\ \citenamefont {Gratieri}}]{Francener:2024euo}%
  \BibitemOpen
  \bibfield  {author} {\bibinfo {author} {\bibfnamefont {R.}~\bibnamefont
  {Francener}}, \bibinfo {author} {\bibfnamefont {V.~P.}\ \bibnamefont
  {Goncalves}}, \ and\ \bibinfo {author} {\bibfnamefont {D.~R.}\ \bibnamefont
  {Gratieri}},\ }\href {\doibase 10.1103/PhysRevD.110.073006} {\bibfield
  {journal} {\bibinfo  {journal} {Phys. Rev. D}\ }\textbf {\bibinfo {volume}
  {110}},\ \bibinfo {pages} {073006} (\bibinfo {year} {2024}{\natexlab{b}})},\
  \Eprint {http://arxiv.org/abs/2408.11736} {arXiv:2408.11736 [hep-ph]}
  \BibitemShut {NoStop}%
\bibitem [{\citenamefont {Huege}\ and\ \citenamefont
  {Kr\"omer}(2024)}]{Huege:2024nic}%
  \BibitemOpen
  \bibfield  {author} {\bibinfo {author} {\bibfnamefont {T.}~\bibnamefont
  {Huege}}\ and\ \bibinfo {author} {\bibfnamefont {O.}~\bibnamefont
  {Kr\"omer}},\ }\href {\doibase 10.1088/1748-0221/19/11/P11022} {\bibfield
  {journal} {\bibinfo  {journal} {JINST}\ }\textbf {\bibinfo {volume} {19}},\
  \bibinfo {pages} {P11022} (\bibinfo {year} {2024})},\ \Eprint
  {http://arxiv.org/abs/2410.22945} {arXiv:2410.22945 [astro-ph.IM]}
  \BibitemShut {NoStop}%
\bibitem [{\citenamefont {Aiello}\ \emph {et~al.}(2025)\citenamefont {Aiello}
  \emph {et~al.}}]{KM3NeT:2025ftj}%
  \BibitemOpen
  \bibfield  {author} {\bibinfo {author} {\bibfnamefont {S.}~\bibnamefont
  {Aiello}} \emph {et~al.} (\bibinfo {collaboration} {KM3NeT}),\ }\href
  {\doibase 10.1007/JHEP07(2025)213} {\bibfield  {journal} {\bibinfo  {journal}
  {JHEP}\ }\textbf {\bibinfo {volume} {07}},\ \bibinfo {pages} {213} (\bibinfo
  {year} {2025})},\ \Eprint {http://arxiv.org/abs/2502.01443} {arXiv:2502.01443
  [hep-ex]} \BibitemShut {NoStop}%
\bibitem [{\citenamefont {Yu}\ \emph {et~al.}(2025)\citenamefont {Yu},
  \citenamefont {Guan}, \citenamefont {Dallaway}, \citenamefont {Rahaman},\
  and\ \citenamefont {Ilic}}]{Yu:2025cyx}%
  \BibitemOpen
  \bibfield  {author} {\bibinfo {author} {\bibfnamefont {X.~Y.}\ \bibnamefont
  {Yu}}, \bibinfo {author} {\bibfnamefont {Z.}~\bibnamefont {Guan}}, \bibinfo
  {author} {\bibfnamefont {W.}~\bibnamefont {Dallaway}}, \bibinfo {author}
  {\bibfnamefont {U.}~\bibnamefont {Rahaman}}, \ and\ \bibinfo {author}
  {\bibfnamefont {N.}~\bibnamefont {Ilic}},\ }\href {\doibase
  10.1007/JHEP07(2025)245} {\bibfield  {journal} {\bibinfo  {journal} {JHEP}\
  }\textbf {\bibinfo {volume} {07}},\ \bibinfo {pages} {245} (\bibinfo {year}
  {2025})},\ \Eprint {http://arxiv.org/abs/2503.16124} {arXiv:2503.16124
  [hep-ph]} \BibitemShut {NoStop}%
\bibitem [{\citenamefont {Ghoshal}\ \emph {et~al.}(2019)\citenamefont
  {Ghoshal}, \citenamefont {Giarnetti},\ and\ \citenamefont
  {Meloni}}]{Ghoshal:2019pab}%
  \BibitemOpen
  \bibfield  {author} {\bibinfo {author} {\bibfnamefont {A.}~\bibnamefont
  {Ghoshal}}, \bibinfo {author} {\bibfnamefont {A.}~\bibnamefont {Giarnetti}},
  \ and\ \bibinfo {author} {\bibfnamefont {D.}~\bibnamefont {Meloni}},\ }\href
  {\doibase 10.1007/JHEP12(2019)126} {\bibfield  {journal} {\bibinfo  {journal}
  {JHEP}\ }\textbf {\bibinfo {volume} {12}},\ \bibinfo {pages} {126} (\bibinfo
  {year} {2019})},\ \Eprint {http://arxiv.org/abs/1906.06212} {arXiv:1906.06212
  [hep-ph]} \BibitemShut {NoStop}%
\bibitem [{\citenamefont {Machado}\ \emph {et~al.}(2020)\citenamefont
  {Machado}, \citenamefont {Schulz},\ and\ \citenamefont
  {Turner}}]{Machado:2020yxl}%
  \BibitemOpen
  \bibfield  {author} {\bibinfo {author} {\bibfnamefont {P.}~\bibnamefont
  {Machado}}, \bibinfo {author} {\bibfnamefont {H.}~\bibnamefont {Schulz}}, \
  and\ \bibinfo {author} {\bibfnamefont {J.}~\bibnamefont {Turner}},\ }\href
  {\doibase 10.1103/PhysRevD.102.053010} {\bibfield  {journal} {\bibinfo
  {journal} {Phys. Rev. D}\ }\textbf {\bibinfo {volume} {102}},\ \bibinfo
  {pages} {053010} (\bibinfo {year} {2020})},\ \Eprint
  {http://arxiv.org/abs/2007.00015} {arXiv:2007.00015 [hep-ph]} \BibitemShut
  {NoStop}%
\bibitem [{\citenamefont {Farzan}\ and\ \citenamefont
  {Tortola}(2018)}]{Farzan:2017xzy}%
  \BibitemOpen
  \bibfield  {author} {\bibinfo {author} {\bibfnamefont {Y.}~\bibnamefont
  {Farzan}}\ and\ \bibinfo {author} {\bibfnamefont {M.}~\bibnamefont
  {Tortola}},\ }\href {\doibase 10.3389/fphy.2018.00010} {\bibfield  {journal}
  {\bibinfo  {journal} {Front. in Phys.}\ }\textbf {\bibinfo {volume} {6}},\
  \bibinfo {pages} {10} (\bibinfo {year} {2018})},\ \Eprint
  {http://arxiv.org/abs/1710.09360} {arXiv:1710.09360 [hep-ph]} \BibitemShut
  {NoStop}%
\bibitem [{\citenamefont {Bhupal~Dev}\ \emph {et~al.}(2019)\citenamefont
  {Bhupal~Dev} \emph {et~al.}}]{Proceedings:2019qno}%
  \BibitemOpen
  \bibfield  {author} {\bibinfo {author} {\bibfnamefont {P.~S.}\ \bibnamefont
  {Bhupal~Dev}} \emph {et~al.},\ }\href {\doibase
  10.21468/SciPostPhysProc.2.001} {\bibfield  {journal} {\bibinfo  {journal}
  {SciPostPhysProc}\ }\textbf {\bibinfo {volume} {2}},\ \bibinfo {pages} {001}
  (\bibinfo {year} {2019})},\ \Eprint {http://arxiv.org/abs/1907.00991}
  {arXiv:1907.00991 [hep-ph]} \BibitemShut {NoStop}%
\bibitem [{\citenamefont {Bischer}\ and\ \citenamefont
  {Rodejohann}(2019)}]{Bischer:2019ttk}%
  \BibitemOpen
  \bibfield  {author} {\bibinfo {author} {\bibfnamefont {I.}~\bibnamefont
  {Bischer}}\ and\ \bibinfo {author} {\bibfnamefont {W.}~\bibnamefont
  {Rodejohann}},\ }\href {\doibase 10.1016/j.nuclphysb.2019.114746} {\bibfield
  {journal} {\bibinfo  {journal} {Nucl. Phys. B}\ }\textbf {\bibinfo {volume}
  {947}},\ \bibinfo {pages} {114746} (\bibinfo {year} {2019})},\ \Eprint
  {http://arxiv.org/abs/1905.08699} {arXiv:1905.08699 [hep-ph]} \BibitemShut
  {NoStop}%
\bibitem [{\citenamefont {Bergmann}\ \emph {et~al.}(2000)\citenamefont
  {Bergmann}, \citenamefont {Grossman},\ and\ \citenamefont
  {Pierce}}]{Bergmann:1999pk}%
  \BibitemOpen
  \bibfield  {author} {\bibinfo {author} {\bibfnamefont {S.}~\bibnamefont
  {Bergmann}}, \bibinfo {author} {\bibfnamefont {Y.}~\bibnamefont {Grossman}},
  \ and\ \bibinfo {author} {\bibfnamefont {D.~M.}\ \bibnamefont {Pierce}},\
  }\href {\doibase 10.1103/PhysRevD.61.053005} {\bibfield  {journal} {\bibinfo
  {journal} {Phys. Rev. D}\ }\textbf {\bibinfo {volume} {61}},\ \bibinfo
  {pages} {053005} (\bibinfo {year} {2000})},\ \Eprint
  {http://arxiv.org/abs/hep-ph/9909390} {arXiv:hep-ph/9909390} \BibitemShut
  {NoStop}%
\bibitem [{\citenamefont {Antusch}\ \emph {et~al.}(2009)\citenamefont
  {Antusch}, \citenamefont {Baumann},\ and\ \citenamefont
  {Fernandez-Martinez}}]{Antusch:2008tz}%
  \BibitemOpen
  \bibfield  {author} {\bibinfo {author} {\bibfnamefont {S.}~\bibnamefont
  {Antusch}}, \bibinfo {author} {\bibfnamefont {J.~P.}\ \bibnamefont
  {Baumann}}, \ and\ \bibinfo {author} {\bibfnamefont {E.}~\bibnamefont
  {Fernandez-Martinez}},\ }\href {\doibase 10.1016/j.nuclphysb.2008.11.018}
  {\bibfield  {journal} {\bibinfo  {journal} {Nucl. Phys. B}\ }\textbf
  {\bibinfo {volume} {810}},\ \bibinfo {pages} {369} (\bibinfo {year}
  {2009})},\ \Eprint {http://arxiv.org/abs/0807.1003} {arXiv:0807.1003
  [hep-ph]} \BibitemShut {NoStop}%
\bibitem [{\citenamefont {Gavela}\ \emph {et~al.}(2009)\citenamefont {Gavela},
  \citenamefont {Hernandez}, \citenamefont {Ota},\ and\ \citenamefont
  {Winter}}]{Gavela:2008ra}%
  \BibitemOpen
  \bibfield  {author} {\bibinfo {author} {\bibfnamefont {M.~B.}\ \bibnamefont
  {Gavela}}, \bibinfo {author} {\bibfnamefont {D.}~\bibnamefont {Hernandez}},
  \bibinfo {author} {\bibfnamefont {T.}~\bibnamefont {Ota}}, \ and\ \bibinfo
  {author} {\bibfnamefont {W.}~\bibnamefont {Winter}},\ }\href {\doibase
  10.1103/PhysRevD.79.013007} {\bibfield  {journal} {\bibinfo  {journal} {Phys.
  Rev. D}\ }\textbf {\bibinfo {volume} {79}},\ \bibinfo {pages} {013007}
  (\bibinfo {year} {2009})},\ \Eprint {http://arxiv.org/abs/0809.3451}
  {arXiv:0809.3451 [hep-ph]} \BibitemShut {NoStop}%
\bibitem [{\citenamefont {Meloni}\ \emph {et~al.}(2010)\citenamefont {Meloni},
  \citenamefont {Ohlsson}, \citenamefont {Winter},\ and\ \citenamefont
  {Zhang}}]{Meloni:2009cg}%
  \BibitemOpen
  \bibfield  {author} {\bibinfo {author} {\bibfnamefont {D.}~\bibnamefont
  {Meloni}}, \bibinfo {author} {\bibfnamefont {T.}~\bibnamefont {Ohlsson}},
  \bibinfo {author} {\bibfnamefont {W.}~\bibnamefont {Winter}}, \ and\ \bibinfo
  {author} {\bibfnamefont {H.}~\bibnamefont {Zhang}},\ }\href {\doibase
  10.1007/JHEP04(2010)041} {\bibfield  {journal} {\bibinfo  {journal} {JHEP}\
  }\textbf {\bibinfo {volume} {04}},\ \bibinfo {pages} {041} (\bibinfo {year}
  {2010})},\ \Eprint {http://arxiv.org/abs/0912.2735} {arXiv:0912.2735
  [hep-ph]} \BibitemShut {NoStop}%
\bibitem [{\citenamefont {Altmannshofer}\ \emph {et~al.}(2019)\citenamefont
  {Altmannshofer}, \citenamefont {Tammaro},\ and\ \citenamefont
  {Zupan}}]{Altmannshofer:2018xyo}%
  \BibitemOpen
  \bibfield  {author} {\bibinfo {author} {\bibfnamefont {W.}~\bibnamefont
  {Altmannshofer}}, \bibinfo {author} {\bibfnamefont {M.}~\bibnamefont
  {Tammaro}}, \ and\ \bibinfo {author} {\bibfnamefont {J.}~\bibnamefont
  {Zupan}},\ }\href {\doibase 10.1007/JHEP11(2021)113} {\bibfield  {journal}
  {\bibinfo  {journal} {JHEP}\ }\textbf {\bibinfo {volume} {09}},\ \bibinfo
  {pages} {083} (\bibinfo {year} {2019})},\ \bibinfo {note} {[Erratum: JHEP 11,
  113 (2021)]},\ \Eprint {http://arxiv.org/abs/1812.02778} {arXiv:1812.02778
  [hep-ph]} \BibitemShut {NoStop}%
\bibitem [{\citenamefont {Falkowski}\ \emph {et~al.}(2020)\citenamefont
  {Falkowski}, \citenamefont {Gonz\'alez-Alonso},\ and\ \citenamefont
  {Tabrizi}}]{Falkowski:2019kfn}%
  \BibitemOpen
  \bibfield  {author} {\bibinfo {author} {\bibfnamefont {A.}~\bibnamefont
  {Falkowski}}, \bibinfo {author} {\bibfnamefont {M.}~\bibnamefont
  {Gonz\'alez-Alonso}}, \ and\ \bibinfo {author} {\bibfnamefont
  {Z.}~\bibnamefont {Tabrizi}},\ }\href {\doibase 10.1007/JHEP11(2020)048}
  {\bibfield  {journal} {\bibinfo  {journal} {JHEP}\ }\textbf {\bibinfo
  {volume} {11}},\ \bibinfo {pages} {048} (\bibinfo {year} {2020})},\ \Eprint
  {http://arxiv.org/abs/1910.02971} {arXiv:1910.02971 [hep-ph]} \BibitemShut
  {NoStop}%
\bibitem [{\citenamefont {Falkowski}\ \emph {et~al.}(2019)\citenamefont
  {Falkowski}, \citenamefont {Gonz\'alez-Alonso},\ and\ \citenamefont
  {Tabrizi}}]{Falkowski:2019xoe}%
  \BibitemOpen
  \bibfield  {author} {\bibinfo {author} {\bibfnamefont {A.}~\bibnamefont
  {Falkowski}}, \bibinfo {author} {\bibfnamefont {M.}~\bibnamefont
  {Gonz\'alez-Alonso}}, \ and\ \bibinfo {author} {\bibfnamefont
  {Z.}~\bibnamefont {Tabrizi}},\ }\href {\doibase 10.1007/JHEP05(2019)173}
  {\bibfield  {journal} {\bibinfo  {journal} {JHEP}\ }\textbf {\bibinfo
  {volume} {05}},\ \bibinfo {pages} {173} (\bibinfo {year} {2019})},\ \Eprint
  {http://arxiv.org/abs/1901.04553} {arXiv:1901.04553 [hep-ph]} \BibitemShut
  {NoStop}%
\bibitem [{\citenamefont {Babu}\ \emph {et~al.}(2020)\citenamefont {Babu},
  \citenamefont {Dev}, \citenamefont {Jana},\ and\ \citenamefont
  {Thapa}}]{Babu:2019mfe}%
  \BibitemOpen
  \bibfield  {author} {\bibinfo {author} {\bibfnamefont {K.~S.}\ \bibnamefont
  {Babu}}, \bibinfo {author} {\bibfnamefont {P.~S.~B.}\ \bibnamefont {Dev}},
  \bibinfo {author} {\bibfnamefont {S.}~\bibnamefont {Jana}}, \ and\ \bibinfo
  {author} {\bibfnamefont {A.}~\bibnamefont {Thapa}},\ }\href {\doibase
  10.1007/JHEP03(2020)006} {\bibfield  {journal} {\bibinfo  {journal} {JHEP}\
  }\textbf {\bibinfo {volume} {03}},\ \bibinfo {pages} {006} (\bibinfo {year}
  {2020})},\ \Eprint {http://arxiv.org/abs/1907.09498} {arXiv:1907.09498
  [hep-ph]} \BibitemShut {NoStop}%
\bibitem [{\citenamefont {Davidson}\ and\ \citenamefont
  {Gorbahn}(2020)}]{Davidson:2019iqh}%
  \BibitemOpen
  \bibfield  {author} {\bibinfo {author} {\bibfnamefont {S.}~\bibnamefont
  {Davidson}}\ and\ \bibinfo {author} {\bibfnamefont {M.}~\bibnamefont
  {Gorbahn}},\ }\href {\doibase 10.1103/PhysRevD.101.015010} {\bibfield
  {journal} {\bibinfo  {journal} {Phys. Rev. D}\ }\textbf {\bibinfo {volume}
  {101}},\ \bibinfo {pages} {015010} (\bibinfo {year} {2020})},\ \Eprint
  {http://arxiv.org/abs/1909.07406} {arXiv:1909.07406 [hep-ph]} \BibitemShut
  {NoStop}%
\bibitem [{\citenamefont {Terol-Calvo}\ \emph {et~al.}(2020)\citenamefont
  {Terol-Calvo}, \citenamefont {T\'ortola},\ and\ \citenamefont
  {Vicente}}]{Terol-Calvo:2019vck}%
  \BibitemOpen
  \bibfield  {author} {\bibinfo {author} {\bibfnamefont {J.}~\bibnamefont
  {Terol-Calvo}}, \bibinfo {author} {\bibfnamefont {M.}~\bibnamefont
  {T\'ortola}}, \ and\ \bibinfo {author} {\bibfnamefont {A.}~\bibnamefont
  {Vicente}},\ }\href {\doibase 10.1103/PhysRevD.101.095010} {\bibfield
  {journal} {\bibinfo  {journal} {Phys. Rev. D}\ }\textbf {\bibinfo {volume}
  {101}},\ \bibinfo {pages} {095010} (\bibinfo {year} {2020})},\ \Eprint
  {http://arxiv.org/abs/1912.09131} {arXiv:1912.09131 [hep-ph]} \BibitemShut
  {NoStop}%
\bibitem [{\citenamefont {Babu}\ \emph {et~al.}(2021)\citenamefont {Babu},
  \citenamefont {Gon\c{c}alves}, \citenamefont {Jana},\ and\ \citenamefont
  {Machado}}]{Babu:2020nna}%
  \BibitemOpen
  \bibfield  {author} {\bibinfo {author} {\bibfnamefont {K.~S.}\ \bibnamefont
  {Babu}}, \bibinfo {author} {\bibfnamefont {D.}~\bibnamefont {Gon\c{c}alves}},
  \bibinfo {author} {\bibfnamefont {S.}~\bibnamefont {Jana}}, \ and\ \bibinfo
  {author} {\bibfnamefont {P.~A.~N.}\ \bibnamefont {Machado}},\ }\href
  {\doibase 10.1016/j.physletb.2021.136131} {\bibfield  {journal} {\bibinfo
  {journal} {Phys. Lett. B}\ }\textbf {\bibinfo {volume} {815}},\ \bibinfo
  {pages} {136131} (\bibinfo {year} {2021})},\ \Eprint
  {http://arxiv.org/abs/2003.03383} {arXiv:2003.03383 [hep-ph]} \BibitemShut
  {NoStop}%
\bibitem [{\citenamefont {Du}\ \emph {et~al.}(2021)\citenamefont {Du},
  \citenamefont {Li}, \citenamefont {Tang}, \citenamefont {Vihonen},\ and\
  \citenamefont {Yu}}]{Du:2020dwr}%
  \BibitemOpen
  \bibfield  {author} {\bibinfo {author} {\bibfnamefont {Y.}~\bibnamefont
  {Du}}, \bibinfo {author} {\bibfnamefont {H.-L.}\ \bibnamefont {Li}}, \bibinfo
  {author} {\bibfnamefont {J.}~\bibnamefont {Tang}}, \bibinfo {author}
  {\bibfnamefont {S.}~\bibnamefont {Vihonen}}, \ and\ \bibinfo {author}
  {\bibfnamefont {J.-H.}\ \bibnamefont {Yu}},\ }\href {\doibase
  10.1007/JHEP03(2021)019} {\bibfield  {journal} {\bibinfo  {journal} {JHEP}\
  }\textbf {\bibinfo {volume} {03}},\ \bibinfo {pages} {019} (\bibinfo {year}
  {2021})},\ \Eprint {http://arxiv.org/abs/2011.14292} {arXiv:2011.14292
  [hep-ph]} \BibitemShut {NoStop}%
\bibitem [{\citenamefont {Falkowski}\ \emph {et~al.}(2021)\citenamefont
  {Falkowski}, \citenamefont {Gonz\'alez-Alonso}, \citenamefont {Kopp},
  \citenamefont {Soreq},\ and\ \citenamefont {Tabrizi}}]{Falkowski:2021bkq}%
  \BibitemOpen
  \bibfield  {author} {\bibinfo {author} {\bibfnamefont {A.}~\bibnamefont
  {Falkowski}}, \bibinfo {author} {\bibfnamefont {M.}~\bibnamefont
  {Gonz\'alez-Alonso}}, \bibinfo {author} {\bibfnamefont {J.}~\bibnamefont
  {Kopp}}, \bibinfo {author} {\bibfnamefont {Y.}~\bibnamefont {Soreq}}, \ and\
  \bibinfo {author} {\bibfnamefont {Z.}~\bibnamefont {Tabrizi}},\ }\href
  {\doibase 10.1007/JHEP10(2021)086} {\bibfield  {journal} {\bibinfo  {journal}
  {JHEP}\ }\textbf {\bibinfo {volume} {10}},\ \bibinfo {pages} {086} (\bibinfo
  {year} {2021})},\ \Eprint {http://arxiv.org/abs/2105.12136} {arXiv:2105.12136
  [hep-ph]} \BibitemShut {NoStop}%
\bibitem [{\citenamefont {Du}\ \emph {et~al.}(2022)\citenamefont {Du},
  \citenamefont {Li}, \citenamefont {Tang}, \citenamefont {Vihonen},\ and\
  \citenamefont {Yu}}]{Du:2021rdg}%
  \BibitemOpen
  \bibfield  {author} {\bibinfo {author} {\bibfnamefont {Y.}~\bibnamefont
  {Du}}, \bibinfo {author} {\bibfnamefont {H.-L.}\ \bibnamefont {Li}}, \bibinfo
  {author} {\bibfnamefont {J.}~\bibnamefont {Tang}}, \bibinfo {author}
  {\bibfnamefont {S.}~\bibnamefont {Vihonen}}, \ and\ \bibinfo {author}
  {\bibfnamefont {J.-H.}\ \bibnamefont {Yu}},\ }\href {\doibase
  10.1103/PhysRevD.105.075022} {\bibfield  {journal} {\bibinfo  {journal}
  {Phys. Rev. D}\ }\textbf {\bibinfo {volume} {105}},\ \bibinfo {pages}
  {075022} (\bibinfo {year} {2022})},\ \Eprint
  {http://arxiv.org/abs/2106.15800} {arXiv:2106.15800 [hep-ph]} \BibitemShut
  {NoStop}%
\bibitem [{\citenamefont {Bres\'o-Pla}\ \emph {et~al.}(2023)\citenamefont
  {Bres\'o-Pla}, \citenamefont {Falkowski}, \citenamefont {Gonz\'alez-Alonso},\
  and\ \citenamefont {Mons\'alvez-Pozo}}]{Breso-Pla:2023tnz}%
  \BibitemOpen
  \bibfield  {author} {\bibinfo {author} {\bibfnamefont {V.}~\bibnamefont
  {Bres\'o-Pla}}, \bibinfo {author} {\bibfnamefont {A.}~\bibnamefont
  {Falkowski}}, \bibinfo {author} {\bibfnamefont {M.}~\bibnamefont
  {Gonz\'alez-Alonso}}, \ and\ \bibinfo {author} {\bibfnamefont
  {K.}~\bibnamefont {Mons\'alvez-Pozo}},\ }\href {\doibase
  10.1007/JHEP05(2023)074} {\bibfield  {journal} {\bibinfo  {journal} {JHEP}\
  }\textbf {\bibinfo {volume} {05}},\ \bibinfo {pages} {074} (\bibinfo {year}
  {2023})},\ \Eprint {http://arxiv.org/abs/2301.07036} {arXiv:2301.07036
  [hep-ph]} \BibitemShut {NoStop}%
\bibitem [{\citenamefont {Kopp}\ \emph {et~al.}(2025)\citenamefont {Kopp},
  \citenamefont {Tabrizi},\ and\ \citenamefont {Urrea}}]{Kopp:2025ffx}%
  \BibitemOpen
  \bibfield  {author} {\bibinfo {author} {\bibfnamefont {J.}~\bibnamefont
  {Kopp}}, \bibinfo {author} {\bibfnamefont {Z.}~\bibnamefont {Tabrizi}}, \
  and\ \bibinfo {author} {\bibfnamefont {S.}~\bibnamefont {Urrea}},\
  }\href@noop {} {\  (\bibinfo {year} {2025})},\ \Eprint
  {http://arxiv.org/abs/2509.21537} {arXiv:2509.21537 [hep-ph]} \BibitemShut
  {NoStop}%
\bibitem [{\citenamefont {de~Blas}\ \emph {et~al.}(2018)\citenamefont
  {de~Blas}, \citenamefont {Criado}, \citenamefont {Perez-Victoria},\ and\
  \citenamefont {Santiago}}]{deBlas:2017xtg}%
  \BibitemOpen
  \bibfield  {author} {\bibinfo {author} {\bibfnamefont {J.}~\bibnamefont
  {de~Blas}}, \bibinfo {author} {\bibfnamefont {J.~C.}\ \bibnamefont {Criado}},
  \bibinfo {author} {\bibfnamefont {M.}~\bibnamefont {Perez-Victoria}}, \ and\
  \bibinfo {author} {\bibfnamefont {J.}~\bibnamefont {Santiago}},\ }\href
  {\doibase 10.1007/JHEP03(2018)109} {\bibfield  {journal} {\bibinfo  {journal}
  {JHEP}\ }\textbf {\bibinfo {volume} {03}},\ \bibinfo {pages} {109} (\bibinfo
  {year} {2018})},\ \Eprint {http://arxiv.org/abs/1711.10391} {arXiv:1711.10391
  [hep-ph]} \BibitemShut {NoStop}%
\bibitem [{\citenamefont {Cherchiglia}\ and\ \citenamefont
  {Santiago}(2024)}]{Cherchiglia:2023aqp}%
  \BibitemOpen
  \bibfield  {author} {\bibinfo {author} {\bibfnamefont {A.}~\bibnamefont
  {Cherchiglia}}\ and\ \bibinfo {author} {\bibfnamefont {J.}~\bibnamefont
  {Santiago}},\ }\href {\doibase 10.1007/JHEP03(2024)018} {\bibfield  {journal}
  {\bibinfo  {journal} {JHEP}\ }\textbf {\bibinfo {volume} {03}},\ \bibinfo
  {pages} {018} (\bibinfo {year} {2024})},\ \Eprint
  {http://arxiv.org/abs/2309.15924} {arXiv:2309.15924 [hep-ph]} \BibitemShut
  {NoStop}%
\bibitem [{\citenamefont {Cherchiglia}(2025)}]{Cherchiglia:2025ufn}%
  \BibitemOpen
  \bibfield  {author} {\bibinfo {author} {\bibfnamefont {A.}~\bibnamefont
  {Cherchiglia}},\ }\href {\doibase 10.3390/universe11070225} {\bibfield
  {journal} {\bibinfo  {journal} {Universe}\ }\textbf {\bibinfo {volume}
  {11}},\ \bibinfo {pages} {225} (\bibinfo {year} {2025})}\BibitemShut
  {NoStop}%
\bibitem [{\citenamefont {Falkowski}\ \emph {et~al.}(2018)\citenamefont
  {Falkowski}, \citenamefont {Grilli~di Cortona},\ and\ \citenamefont
  {Tabrizi}}]{Falkowski:2018dmy}%
  \BibitemOpen
  \bibfield  {author} {\bibinfo {author} {\bibfnamefont {A.}~\bibnamefont
  {Falkowski}}, \bibinfo {author} {\bibfnamefont {G.}~\bibnamefont {Grilli~di
  Cortona}}, \ and\ \bibinfo {author} {\bibfnamefont {Z.}~\bibnamefont
  {Tabrizi}},\ }\href {\doibase 10.1007/JHEP04(2018)101} {\bibfield  {journal}
  {\bibinfo  {journal} {JHEP}\ }\textbf {\bibinfo {volume} {04}},\ \bibinfo
  {pages} {101} (\bibinfo {year} {2018})},\ \Eprint
  {http://arxiv.org/abs/1802.08296} {arXiv:1802.08296 [hep-ph]} \BibitemShut
  {NoStop}%
\bibitem [{\citenamefont {Kopp}\ \emph {et~al.}(2024)\citenamefont {Kopp},
  \citenamefont {Rocco},\ and\ \citenamefont {Tabrizi}}]{Kopp:2024yvh}%
  \BibitemOpen
  \bibfield  {author} {\bibinfo {author} {\bibfnamefont {J.}~\bibnamefont
  {Kopp}}, \bibinfo {author} {\bibfnamefont {N.}~\bibnamefont {Rocco}}, \ and\
  \bibinfo {author} {\bibfnamefont {Z.}~\bibnamefont {Tabrizi}},\ }\href
  {\doibase 10.1007/JHEP08(2024)187} {\bibfield  {journal} {\bibinfo  {journal}
  {JHEP}\ }\textbf {\bibinfo {volume} {08}},\ \bibinfo {pages} {187} (\bibinfo
  {year} {2024})},\ \Eprint {http://arxiv.org/abs/2401.07902} {arXiv:2401.07902
  [hep-ph]} \BibitemShut {NoStop}%
\bibitem [{\citenamefont {Maki}\ \emph {et~al.}(1962)\citenamefont {Maki},
  \citenamefont {Nakagawa},\ and\ \citenamefont {Sakata}}]{Maki:1962mu}%
  \BibitemOpen
  \bibfield  {author} {\bibinfo {author} {\bibfnamefont {Z.}~\bibnamefont
  {Maki}}, \bibinfo {author} {\bibfnamefont {M.}~\bibnamefont {Nakagawa}}, \
  and\ \bibinfo {author} {\bibfnamefont {S.}~\bibnamefont {Sakata}},\ }\href
  {\doibase 10.1143/PTP.28.870} {\bibfield  {journal} {\bibinfo  {journal}
  {Prog. Theor. Phys.}\ }\textbf {\bibinfo {volume} {28}},\ \bibinfo {pages}
  {870} (\bibinfo {year} {1962})}\BibitemShut {NoStop}%
\bibitem [{\citenamefont {Pontecorvo}(1957)}]{Pontecorvo:1957cp}%
  \BibitemOpen
  \bibfield  {author} {\bibinfo {author} {\bibfnamefont {B.}~\bibnamefont
  {Pontecorvo}},\ }\href@noop {} {\bibfield  {journal} {\bibinfo  {journal}
  {Sov. Phys. JETP}\ }\textbf {\bibinfo {volume} {6}},\ \bibinfo {pages} {429}
  (\bibinfo {year} {1957})},\ \bibinfo {note} {[Zh. Eksp. Teor.
  Fiz.33,549(1957)]}\BibitemShut {NoStop}%
\bibitem [{\citenamefont {Cherchiglia}\ \emph {et~al.}(2023)\citenamefont
  {Cherchiglia}, \citenamefont {Pasquini}, \citenamefont {Peres}, \citenamefont
  {Rodrigues}, \citenamefont {Rossi},\ and\ \citenamefont
  {Souza}}]{Cherchiglia:2023ojf}%
  \BibitemOpen
  \bibfield  {author} {\bibinfo {author} {\bibfnamefont {A.}~\bibnamefont
  {Cherchiglia}}, \bibinfo {author} {\bibfnamefont {P.}~\bibnamefont
  {Pasquini}}, \bibinfo {author} {\bibfnamefont {O.~L.~G.}\ \bibnamefont
  {Peres}}, \bibinfo {author} {\bibfnamefont {F.~F.}\ \bibnamefont
  {Rodrigues}}, \bibinfo {author} {\bibfnamefont {R.~R.}\ \bibnamefont
  {Rossi}}, \ and\ \bibinfo {author} {\bibfnamefont {E.~S.}\ \bibnamefont
  {Souza}},\ }\href {\doibase 10.1103/55qm-zbhv} {\  (\bibinfo {year} {2023}),\
  10.1103/55qm-zbhv},\ \Eprint {http://arxiv.org/abs/2310.18401}
  {arXiv:2310.18401 [hep-ph]} \BibitemShut {NoStop}%
\bibitem [{\citenamefont {Navas}\ \emph {et~al.}(2024)\citenamefont {Navas}
  \emph {et~al.}}]{ParticleDataGroup:2024cfk}%
  \BibitemOpen
  \bibfield  {author} {\bibinfo {author} {\bibfnamefont {S.}~\bibnamefont
  {Navas}} \emph {et~al.} (\bibinfo {collaboration} {Particle Data Group}),\
  }\href {\doibase 10.1103/PhysRevD.110.030001} {\bibfield  {journal} {\bibinfo
   {journal} {Phys. Rev. D}\ }\textbf {\bibinfo {volume} {110}},\ \bibinfo
  {pages} {030001} (\bibinfo {year} {2024})}\BibitemShut {NoStop}%
\bibitem [{\citenamefont {Esteban}\ \emph {et~al.}(2024)\citenamefont
  {Esteban}, \citenamefont {Gonzalez-Garcia}, \citenamefont {Maltoni},
  \citenamefont {Martinez-Soler}, \citenamefont {Pinheiro},\ and\ \citenamefont
  {Schwetz}}]{Esteban:2024eli}%
  \BibitemOpen
  \bibfield  {author} {\bibinfo {author} {\bibfnamefont {I.}~\bibnamefont
  {Esteban}}, \bibinfo {author} {\bibfnamefont {M.~C.}\ \bibnamefont
  {Gonzalez-Garcia}}, \bibinfo {author} {\bibfnamefont {M.}~\bibnamefont
  {Maltoni}}, \bibinfo {author} {\bibfnamefont {I.}~\bibnamefont
  {Martinez-Soler}}, \bibinfo {author} {\bibfnamefont {J.~a.~P.}\ \bibnamefont
  {Pinheiro}}, \ and\ \bibinfo {author} {\bibfnamefont {T.}~\bibnamefont
  {Schwetz}},\ }\href {\doibase 10.1007/JHEP12(2024)216} {\bibfield  {journal}
  {\bibinfo  {journal} {JHEP}\ }\textbf {\bibinfo {volume} {12}},\ \bibinfo
  {pages} {216} (\bibinfo {year} {2024})},\ \Eprint
  {http://arxiv.org/abs/2410.05380} {arXiv:2410.05380 [hep-ph]} \BibitemShut
  {NoStop}%
\bibitem [{\citenamefont {Asano}\ and\ \citenamefont
  {Minakata}(2011)}]{Asano:2011nj}%
  \BibitemOpen
  \bibfield  {author} {\bibinfo {author} {\bibfnamefont {K.}~\bibnamefont
  {Asano}}\ and\ \bibinfo {author} {\bibfnamefont {H.}~\bibnamefont
  {Minakata}},\ }\href {\doibase 10.1007/JHEP06(2011)022} {\bibfield  {journal}
  {\bibinfo  {journal} {JHEP}\ }\textbf {\bibinfo {volume} {06}},\ \bibinfo
  {pages} {022} (\bibinfo {year} {2011})},\ \Eprint
  {http://arxiv.org/abs/1103.4387} {arXiv:1103.4387 [hep-ph]} \BibitemShut
  {NoStop}%
\bibitem [{\citenamefont {Abi}\ \emph {et~al.}(2021)\citenamefont {Abi} \emph
  {et~al.}}]{DUNE:2021cuw}%
  \BibitemOpen
  \bibfield  {author} {\bibinfo {author} {\bibfnamefont {B.}~\bibnamefont
  {Abi}} \emph {et~al.} (\bibinfo {collaboration} {DUNE}),\ }\href@noop {} {\
  (\bibinfo {year} {2021})},\ \Eprint {http://arxiv.org/abs/2103.04797}
  {arXiv:2103.04797 [hep-ex]} \BibitemShut {NoStop}%
\bibitem [{\citenamefont {Abi}\ \emph {et~al.}(2020{\natexlab{b}})\citenamefont
  {Abi} \emph {et~al.}}]{DUNE:2020jqi}%
  \BibitemOpen
  \bibfield  {author} {\bibinfo {author} {\bibfnamefont {B.}~\bibnamefont
  {Abi}} \emph {et~al.} (\bibinfo {collaboration} {DUNE}),\ }\href {\doibase
  10.1140/epjc/s10052-020-08456-z} {\bibfield  {journal} {\bibinfo  {journal}
  {Eur. Phys. J. C}\ }\textbf {\bibinfo {volume} {80}},\ \bibinfo {pages} {978}
  (\bibinfo {year} {2020}{\natexlab{b}})},\ \Eprint
  {http://arxiv.org/abs/2006.16043} {arXiv:2006.16043 [hep-ex]} \BibitemShut
  {NoStop}%
\bibitem [{\citenamefont {Huber}\ \emph {et~al.}(2005)\citenamefont {Huber},
  \citenamefont {Lindner},\ and\ \citenamefont {Winter}}]{Huber:2004ka}%
  \BibitemOpen
  \bibfield  {author} {\bibinfo {author} {\bibfnamefont {P.}~\bibnamefont
  {Huber}}, \bibinfo {author} {\bibfnamefont {M.}~\bibnamefont {Lindner}}, \
  and\ \bibinfo {author} {\bibfnamefont {W.}~\bibnamefont {Winter}},\ }\href
  {\doibase 10.1016/j.cpc.2005.01.003} {\bibfield  {journal} {\bibinfo
  {journal} {Comput. Phys. Commun.}\ }\textbf {\bibinfo {volume} {167}},\
  \bibinfo {pages} {195} (\bibinfo {year} {2005})},\ \Eprint
  {http://arxiv.org/abs/hep-ph/0407333} {arXiv:hep-ph/0407333} \BibitemShut
  {NoStop}%
\bibitem [{\citenamefont {Huber}\ \emph {et~al.}(2007)\citenamefont {Huber},
  \citenamefont {Kopp}, \citenamefont {Lindner}, \citenamefont {Rolinec},\ and\
  \citenamefont {Winter}}]{Huber:2007ji}%
  \BibitemOpen
  \bibfield  {author} {\bibinfo {author} {\bibfnamefont {P.}~\bibnamefont
  {Huber}}, \bibinfo {author} {\bibfnamefont {J.}~\bibnamefont {Kopp}},
  \bibinfo {author} {\bibfnamefont {M.}~\bibnamefont {Lindner}}, \bibinfo
  {author} {\bibfnamefont {M.}~\bibnamefont {Rolinec}}, \ and\ \bibinfo
  {author} {\bibfnamefont {W.}~\bibnamefont {Winter}},\ }\href {\doibase
  10.1016/j.cpc.2007.05.004} {\bibfield  {journal} {\bibinfo  {journal}
  {Comput. Phys. Commun.}\ }\textbf {\bibinfo {volume} {177}},\ \bibinfo
  {pages} {432} (\bibinfo {year} {2007})},\ \Eprint
  {http://arxiv.org/abs/hep-ph/0701187} {arXiv:hep-ph/0701187} \BibitemShut
  {NoStop}%
\bibitem [{\citenamefont {Blennow}\ and\ \citenamefont
  {Fernandez-Martinez}(2010)}]{Blennow:2009pk}%
  \BibitemOpen
  \bibfield  {author} {\bibinfo {author} {\bibfnamefont {M.}~\bibnamefont
  {Blennow}}\ and\ \bibinfo {author} {\bibfnamefont {E.}~\bibnamefont
  {Fernandez-Martinez}},\ }\href {\doibase 10.1016/j.cpc.2009.09.014}
  {\bibfield  {journal} {\bibinfo  {journal} {Comput. Phys. Commun.}\ }\textbf
  {\bibinfo {volume} {181}},\ \bibinfo {pages} {227} (\bibinfo {year}
  {2010})},\ \Eprint {http://arxiv.org/abs/0903.3985} {arXiv:0903.3985
  [hep-ph]} \BibitemShut {NoStop}%
\bibitem [{\citenamefont {Guzzo}\ \emph {et~al.}(2023)\citenamefont {Guzzo},
  \citenamefont {Leite}, \citenamefont {Novelo}, \citenamefont {Peres},\ and\
  \citenamefont {Pleitez}}]{Guzzo:2023ayo}%
  \BibitemOpen
  \bibfield  {author} {\bibinfo {author} {\bibfnamefont {M.~M.}\ \bibnamefont
  {Guzzo}}, \bibinfo {author} {\bibfnamefont {L.~J.~F.}\ \bibnamefont {Leite}},
  \bibinfo {author} {\bibfnamefont {S.~W.~P.}\ \bibnamefont {Novelo}}, \bibinfo
  {author} {\bibfnamefont {O.~L.~G.}\ \bibnamefont {Peres}}, \ and\ \bibinfo
  {author} {\bibfnamefont {V.}~\bibnamefont {Pleitez}},\ }\href {\doibase
  10.1103/PhysRevD.107.095037} {\bibfield  {journal} {\bibinfo  {journal}
  {Phys. Rev. D}\ }\textbf {\bibinfo {volume} {107}},\ \bibinfo {pages}
  {095037} (\bibinfo {year} {2023})},\ \Eprint
  {http://arxiv.org/abs/2102.13118} {arXiv:2102.13118 [hep-ph]} \BibitemShut
  {NoStop}%
\bibitem [{\citenamefont {Cirigliano}\ and\ \citenamefont
  {Rosell}(2007)}]{Cirigliano:2007xi}%
  \BibitemOpen
  \bibfield  {author} {\bibinfo {author} {\bibfnamefont {V.}~\bibnamefont
  {Cirigliano}}\ and\ \bibinfo {author} {\bibfnamefont {I.}~\bibnamefont
  {Rosell}},\ }\href {\doibase 10.1103/PhysRevLett.99.231801} {\bibfield
  {journal} {\bibinfo  {journal} {Phys. Rev. Lett.}\ }\textbf {\bibinfo
  {volume} {99}},\ \bibinfo {pages} {231801} (\bibinfo {year} {2007})},\
  \Eprint {http://arxiv.org/abs/0707.3439} {arXiv:0707.3439 [hep-ph]}
  \BibitemShut {NoStop}%
\bibitem [{\citenamefont {Bryman}\ \emph {et~al.}(2011)\citenamefont {Bryman},
  \citenamefont {Marciano}, \citenamefont {Tschirhart},\ and\ \citenamefont
  {Yamanaka}}]{Bryman:2011zz}%
  \BibitemOpen
  \bibfield  {author} {\bibinfo {author} {\bibfnamefont {D.}~\bibnamefont
  {Bryman}}, \bibinfo {author} {\bibfnamefont {W.~J.}\ \bibnamefont
  {Marciano}}, \bibinfo {author} {\bibfnamefont {R.}~\bibnamefont
  {Tschirhart}}, \ and\ \bibinfo {author} {\bibfnamefont {T.}~\bibnamefont
  {Yamanaka}},\ }\href {\doibase 10.1146/annurev-nucl-102010-130431} {\bibfield
   {journal} {\bibinfo  {journal} {Ann. Rev. Nucl. Part. Sci.}\ }\textbf
  {\bibinfo {volume} {61}},\ \bibinfo {pages} {331} (\bibinfo {year}
  {2011})}\BibitemShut {NoStop}%
\bibitem [{\citenamefont {Aguilar-Arevalo}\ \emph {et~al.}(2015)\citenamefont
  {Aguilar-Arevalo} \emph {et~al.}}]{PiENu:2015seu}%
  \BibitemOpen
  \bibfield  {author} {\bibinfo {author} {\bibfnamefont {A.}~\bibnamefont
  {Aguilar-Arevalo}} \emph {et~al.} (\bibinfo {collaboration} {PiENu}),\ }\href
  {\doibase 10.1103/PhysRevLett.115.071801} {\bibfield  {journal} {\bibinfo
  {journal} {Phys. Rev. Lett.}\ }\textbf {\bibinfo {volume} {115}},\ \bibinfo
  {pages} {071801} (\bibinfo {year} {2015})},\ \Eprint
  {http://arxiv.org/abs/1506.05845} {arXiv:1506.05845 [hep-ex]} \BibitemShut
  {NoStop}%
\bibitem [{\citenamefont {Baker}\ and\ \citenamefont
  {Cousins}(1984)}]{Baker:1983tu}%
  \BibitemOpen
  \bibfield  {author} {\bibinfo {author} {\bibfnamefont {S.}~\bibnamefont
  {Baker}}\ and\ \bibinfo {author} {\bibfnamefont {R.~D.}\ \bibnamefont
  {Cousins}},\ }\href {\doibase 10.1016/0167-5087(84)90016-4} {\bibfield
  {journal} {\bibinfo  {journal} {Nucl. Instrum. Meth.}\ }\textbf {\bibinfo
  {volume} {221}},\ \bibinfo {pages} {437} (\bibinfo {year}
  {1984})}\BibitemShut {NoStop}%
\bibitem [{\citenamefont {Almeida~Jr.}\ \emph {et~al.}(2000)\citenamefont
  {Almeida~Jr.}, \citenamefont {Barbi},\ and\ \citenamefont
  {do~Vale}}]{Almeida:1999ie}%
  \BibitemOpen
  \bibfield  {author} {\bibinfo {author} {\bibfnamefont {F.~M.~L.}\
  \bibnamefont {Almeida~Jr.}}, \bibinfo {author} {\bibfnamefont
  {M.}~\bibnamefont {Barbi}}, \ and\ \bibinfo {author} {\bibfnamefont
  {M.~A.~B.}\ \bibnamefont {do~Vale}},\ }\href {\doibase
  10.1016/S0168-9002(99)01466-7} {\bibfield  {journal} {\bibinfo  {journal}
  {Nucl. Instrum. Meth. A}\ }\textbf {\bibinfo {volume} {449}},\ \bibinfo
  {pages} {383} (\bibinfo {year} {2000})},\ \Eprint
  {http://arxiv.org/abs/hep-ex/9911042} {arXiv:hep-ex/9911042} \BibitemShut
  {NoStop}%
\bibitem [{\citenamefont {Dziewonski}\ and\ \citenamefont
  {Anderson}(1981)}]{Dziewonski:1981xy}%
  \BibitemOpen
  \bibfield  {author} {\bibinfo {author} {\bibfnamefont {A.~M.}\ \bibnamefont
  {Dziewonski}}\ and\ \bibinfo {author} {\bibfnamefont {D.~L.}\ \bibnamefont
  {Anderson}},\ }\href {\doibase 10.1016/0031-9201(81)90046-7} {\bibfield
  {journal} {\bibinfo  {journal} {Phys. Earth Planet. Interiors}\ }\textbf
  {\bibinfo {volume} {25}},\ \bibinfo {pages} {297} (\bibinfo {year}
  {1981})}\BibitemShut {NoStop}%
\bibitem [{\citenamefont {Peres}(2025)}]{santafe-conference}%
  \BibitemOpen
  \bibfield  {author} {\bibinfo {author} {\bibfnamefont {O.~L.~G.}\
  \bibnamefont {Peres}},\ }\href
  {https://indico.nevis.columbia.edu/event/6/contributions/40/} {\enquote
  {\bibinfo {title} {Cc nsi in short baseline experiments},}\ } (\bibinfo
  {year} {2025}),\ \bibinfo {note} {2nd Short-Baseline Experiment-Theory
  Workshop}\BibitemShut {NoStop}%
\end{thebibliography}%

\end{document}